\documentclass[superscriptaddress, twocolumn, nofootinbib,aps]{revtex4-2}

\usepackage{amssymb,amsmath,amsfonts}
\usepackage{latexsym}
\usepackage{graphicx}
\usepackage{caption}
\usepackage{subcaption}
\usepackage{enumerate}
\usepackage{float}
\usepackage{bm}
\usepackage{mathrsfs}
\usepackage{color}
\usepackage{graphics}
\usepackage{epsfig}
\usepackage{bbm}  
\usepackage{soul}   
\usepackage{scalerel,stackengine}
\stackMath
\usepackage{color}
\usepackage{xcolor}
\usepackage[breaklinks,colorlinks,backref]{hyperref}
\hypersetup{
    colorlinks, 
    linktoc=all, 
    linkcolor=red,  
  citecolor=hyptxt,
  urlcolor=blue}
  \hypersetup{
  citebordercolor=,
  filebordercolor=green,
  linkbordercolor=blue
}
\definecolor{hyptxt}{rgb}{0.7, 0.4, 0.9}

\newcommand\reallywidecheck[1]{%
\savestack{\tmpbox}{\stretchto{%
  \scaleto{%
    \scalerel*[\widthof{\ensuremath{#1}}]{\kern-.6pt\bigwedge\kern-.6pt}%
    {\rule[-\textheight/2]{1ex}{\textheight}}
  }{\textheight}%
}{0.5ex}}%
\stackon[1pt]{#1}{\scalebox{-1}{\tmpbox}}%
}

\newcounter{mnotecount}[section]

\begin{document}

\newcommand{\dR}{\mathbb R}
\newcommand{\dC}{\mathbb C}
\newcommand{\dZ}{\mathbb Z}
\newcommand{\id}{\mathbb I}
\newtheorem{theorem}{Theorem}
\newcommand{\ud}{\mathrm{d}}
\newcommand{\mfn}{\mathfrak{n}}

\def\calP{\mathcal{P}}
\def\calT{\mathcal{T}}
\def\CP{M}
\def\CC{\mathcal{C}}
\def\BCP{\boldsymbol{M}}
\def\BCC{\boldsymbol{\mathcal{C}}}
\def\BCH{\boldsymbol{\mathcal{H}}}
\def\BRF{\boldsymbol{\mathrm{F}}}
\def\BP{\boldsymbol{\Psi}}
\def\BLI{\boldsymbol{\Lambda}_{P}}
\def\BLC{\boldsymbol{\Lambda}_{C}}
\def\tint{{\textstyle\int}}
\def\lg{\langle }
\def\rg{\rangle }
\def\mg{{ \mathrm{\bf H}}}
\def\sh{\underset{\mbox{\footnotesize s}}{|}}
\def\dk{\underset{\mbox{\footnotesize f}}{|}}
\def\adg{a^{\dag}}
\def\vp{\varphi}
\def\vo{\varomega}
\def\ve{\varepsilon}
\def\bz{\bar{z}}
\def\deq{\stackrel{\mathrm{def}}{=}}
\def\vr{\hat{r}}
\def\half{\frac{1}{2}}
\def\lu{\mbox{\large 1}}
\def\ud{\mathrm{d}}
\def\mfn{\mathfrak{n}}
\def\calH{{\cal H }}
\def\K{{\mathcal K }}
\def\H{\mathbb{H}}
\def\R{\mathbb{R}}
\def\N{\mathbb{N}}
\def\C{\mathbb{C}}
\def\Z {\mathbb{Z}}
\def\I{\mathbb{I}}
\def\vep{\varepsilon}
\def\FJ{\mathfrak{J}}
\def\RJJ{\mathcal{R}_{\mathfrak{J}}}
\def\TFJ{\tilde{\mathfrak{J}}}
\def\TE{\tilde{E}}
\def\vep{\varepsilon}
\def\vap{\varpi}
\def\FJ{\mathfrak{J}}
\def\RJJ{\mathcal{R}_{\mathfrak{J}}}
\def\TFJ{\tilde{\mathfrak{J}}}
\def\QE{\mathscr{E}}
\def\TE{\tilde{E}}
\def\TQE{\tilde{\QE}}
\def\lu{\mbox{\large 1}}
\def\calP{\mathcal{P}}
\def\calT{\mathcal{T}}
\def\CP{M}
\def\btr{\blacktriangleright}
\def\ii{\mathrm{i}}
\def\sfh{\mathsf{h}}
\def\sfH{\mathsf{H}}
\def\sfK{\mathsf{K}}
\def\sfM{\mathsf{M}}
\def\sfMps_{\sfM_{\sigma}^{\vap}}
\def\sfC{\mathsf{C}}
\def\sfP{\mathsf{P}}
\def\sfMv{\mathsf{M}^{\vap}}
\def\cMv{\mathcal{M}^{\vap}}
\def\sft{\mathsf{t}}
\def\sfMa{\sfM^{a\mathcal{W}}}
\def\vapa{\vap_{a\mathcal{W}}}
\def\bsft{\mathbf{ \mathsf{t}}}
\def\sfd{\mathsf{d}}
\def\bsfd{\mathbf{ \mathsf{d}}}
\def\mfd{\mathfrak{D}}
\def\mfF{\mathfrak{F}}
\def\mfG{\mathfrak{G}}
\def\mfA{\mathfrak{A}}
\def\mFs{\mathfrak{F_s}}
\def\mfs{\mathfrak{f_s}}
\def\omfs{\overline{\mathfrak{f_s}}}
\def\cdm{\mathsf{C}_{\mathrm{DM}}}
\def\mfn{\mathfrak{n}}
\def\vr{\vec{\mathbf{r}}}
\def\vrp{\vec{\mathbf{r}^{\prime}}}
\def\vrpp{\vec{\mathbf{r}^{\prime\prime}}}
\def\vz{\vec{\mathbf{0}}}
\def\ud{\mathrm{d}}
\def\lg{\langle }
\def\rg{\rangle }
\def\arcosh{\textrm{arcosh}}
\def\bsb{\boldsymbol{\beta}}
\def\wbsb{\widehat\bsb}
\def\hbp{\hat\beta_+}
\def\hbm{\hat\beta_-}
\def\hq{\hat q}
\def\hp{\hat p}
\def\ii{\mathrm{i}}
\def\bu{\mathbbm{1}}

\title[]{Can a quantum mixmaster universe undergo a spontaneous inflationary phase?}
\author{Herv\'e Bergeron}

\address{ISMO, UMR 8214 CNRS, Univ Paris-Saclay,  91405 Orsay CEDEX, France}
\email{herve.bergeron@universite-paris-saclay.fr}

\author{Jaime de Cabo Martin}
\address{National Centre for Nuclear Research, Pasteura 7, 02-093 Warsaw, Poland,}
\email{jaime.decabomartin@ncbj.gov.pl}

\author{Jean-Pierre Gazeau}
\address{Universit\'e Paris Cit\'e, CNRS, Astroparticule et Cosmologie, 75013 Paris, France } 
\email{gazeau@apc.in2p3.fr}

\author{Przemys\l aw Ma\l kiewicz}
\address{National Centre for Nuclear Research, Pasteura 7, 02-093 Warsaw, Poland}
\email{Przemyslaw.Malkiewicz@ncbj.gov.pl}
\date{\today}
\begin{abstract}
We study a semiclassical model of the mixmaster universe. We first derive the quantum model and then introduce its semiclassical approximation. We employ a general integral quantization method that respects the symmetries of the model given by the affine and the Weyl-Heisenberg groups, and can produce a wide class of quantum models. The semiclassical approximation is based on the coherent states. The semiclassical dynamics is complex and cannot be solved by analytical methods. We focus on a key qualitative feature of the dynamics, namely, we investigate whether the primordial anisotropic universe can undergo a spontaneous inflationary phase driven by the anisotropic energy combined with semiclassical corrections. The answer to this question provides a useful perspective on the inflationary paradigm as well as on alternative bouncing models. 
\end{abstract}
\maketitle

\section{Introduction}

Cosmic inflation generated by a single scalar field called inflaton is the current paradigm for the theory of the origin of primordial structure in the Universe \cite{intro}. Nevertheless, it still faces some problems such as restrictions on the initial condition, the fine-tuning of the inflationary potential or the multiverse problem (see e.g. \cite{infl1} and also \cite{infl2,infl3}). Alternative models for the primordial universe are bouncing cosmologies in which the generation of primordial structure occurs during the contraction and the bounce that stops contraction and initiates the present expansion. Unfortunately, the simplest bouncing models tend to generate blue-tilted spectrum \cite{Peter:2006id, Peter:2006hx, Martin2004a, Malkiewicz:2020fvy,Malk2023} for primordial perturbations contrary to the observational evidence. We note that the both types of models share a very restrictive assumption of a slightly perturbed isotropic and homogenous universe. However, substantial amounts of inhomogeneities and anisotropies could play an important role in the primordial universe: on one hand, they could hinder the cosmic inflation driven by inflaton while, on the other hand, they themselves could spontaneously generate an accelerated expansion phase. The latter possibility was discussed, e.g., in \cite{thomasbuchert}. Unfortunately, a nonperturbative investigation into the inhomogeneous primordial universe remains a very challenging problem. Nevertheless, a less demanding question, though still utterly important, of whether a sufficient amount of anisotropy in the primordial universe could spontaneously generate a cosmic inflation turns out to be tractable. To our best knowledge, this question has never been studied apart from a few related works that we mention below. 

In this work we study the anisotropic Bianchi Type IX model of the universe, also known as the mixmaster universe. We quantize the Bianchi IX model and introduce a semiclassical framework in which its dynamics is more accessible, though far from trivial. The employed quantization procedure encompasses many quantization ambiguities, which makes our study more general. It respects the symmetries of the phase space of the Bianchi IX model and produces a self-adjoint representation of relevant observables such as the Hamiltonian. The main outcome of the employed quantization is the resolution of the big-bang singularity via a bouncing dynamics as well as a modification to the anisotropy potential. The semiclassical framework is derived with the use of coherent states that also respect the existing symmetries. The latter are given by the product of the  Weyl-Heisenberg and the affine group. The semiclassical phase space is showed to exhibit a generic bounce replacing the big-bang and big-crunch singularities. This part of our paper is largely an improved and self-contained presentation of our previous results that interested reader can also find in \cite{berczugamal20,berczgamapie15A,berczgamapie15B,berczgama16A,berczgama16B}.

In the main part of the work we deal with the role of anisotropy in the semiclassical dynamics close to the bounce. The classical dynamics of the mixmaster universe is widely known to be very complex. The employed quantization combined with our semiclassical framework produce a model of similar, if not higher, complexity. Therefore, we address our specific question about the mixmaster dynamics in qualitative terms, which permits to avoid the mathematical and numerical difficulties of finding the full solution. Our previous result \cite{berczgama16A,berczgama16B} suggests that as the universe emerges from the bounce the anisotropy continues to be strongly excited and forcing the isotropic geometry of the universe to expand in an accelerated way and for a long period of time. In other words, the quantum mixmaster universe seems to spontaneously generate an inflationary phase. The use of the words "suggests" and "seems" is fair as for deriving that result we used a crude approximation to the anisotropy potential, though the analyzed model was fully quantum. In the present work we resolve this issue in a semiclassical framework without making any approximation to the anisotropy potential.

The significance of the investigated issue is clear. The existence of a robust inflationary phase in a bouncing anisotropic model could provide a serious challenge to the hypothesis of inflaton and its paramount role in the primordial evolution. On the other hand, the nonexistence of such a phase in our model should in principle strengthen the existing arguments in favor of inflaton as another attempt at challenging its exceptional status fails. 

The plan of our paper is as follows: In Sec \ref{prelim} we briefly recall the Hamiltonian formalism for the Bianchi IX model. In Sec \ref{quantsmod} we explain in a compact but self-contained manner the covariant quantization method and the semiclassical framework with application to the studied model. In Sec \ref{semiclassdyn} we discuss the main features of the semiclassical dynamics and illustrate them with a few numerical examples. In Sec \ref{accelerated} we obtain the answer to the title question. The discussion of the result and conclusions are found in Sec \ref{disc}.

\section{Classical model}
\label{prelim}
We first recall the Hamiltonian formulation of the Bianchi type IX model. We assume the following line element:
\begin{equation}
\ud s^2= -{\cal N}^2\ud\tau^2+\sum_ia_i^2(\omega^i)^2\, ,
\end{equation}
where {$\ud \omega_i=\frac{1}{2} \frak{n}\varepsilon_{i}^{\, jk}\omega_j \wedge \omega_k$, ${\cal N}(\tau)$ and $a_i(\tau)$ are positive-valued functions of time. The Hamiltonian constraint of this spacetime model expressed in the Misner variables $(\Omega,p_{\Omega},\bsb, \mathbf{p})\in\mathbb{R}^6$ reads \cite{misner69}:
\begin{align}\label{con}\begin{split}
&\mathrm{C}=\frac{{\cal N}e^{-3\Omega}}{24}\left(\frac{\mathcal{V}_0}{2\kappa}\right)\\
&\times\left(\left(\frac{2\kappa}{\mathcal{V}_0}\right)^2[-p_{\Omega}^2+\mathbf{ p}^2]+36\frak{n}^2e^{4\Omega}[V(\bsb)-1]\right),\end{split}
\end{align}
where $\bsb:=(\beta_{+},\beta_-)\in \R^2$ and $\mathbf{p}:= (p_+,p_-)\in \R^2$ are canonically conjugate variables, $\mathcal{V}_0=16\pi^2/\frak{n}^3$ is the coordinate volume of the spatial section, $\kappa=8\pi G$ is the gravitational constant. The lapse function ${\cal N}$ plays the role of Lagrange multiplier. {In what follows we set $\frak{n}=1$, which implies $\mathcal{V}_0=16\pi^2$. We also set the physical dimensions in such a way that $2\kappa=\mathcal{V}_0$, which implies $G=\pi$.} The gravitational Hamiltonian $\mathrm{C}$ resembles the Hamiltonian of a particle in the 3D Minkowski spacetime moving in a time-dependent potential. The Misner variables have the following cosmological interpretation: 
\begin{equation}\Omega=\frac13\ln a_1a_2a_3,~~\beta_+=\frac16\ln\frac{a_1a_2}{a_3^2},~~\beta_-=\frac{1}{2\sqrt{3}}\ln\frac{a_1}{a_2} . \end{equation} 
Clearly, the variable $\Omega$ describes the isotropic part of geometry, whereas $\beta_{\pm}$ describe the distortions to isotropy and are called the anisotropic variables. The potential that drives the motion of the geometry originates from the spatial curvature, and reads
\begin{equation}\label{b9pot}
V(\bsb) = \frac{e^{4\beta_+}}{3} \left[\left(2\cosh(2\sqrt{3}\beta_-)-e^{- 6\beta_+}
\right)^2-4\right] +  1 \,.
\end{equation}

Following our previous papers \cite{berczgamapie15A,berczgamapie15B,berczgama16B,berczgama16A,berczugamal17,ewa2018} we redefine the isotropic variables as follows:
\begin{align}
\label{qp}
q=e^{\frac{3}{2}\Omega},~~p=\frac{2}{3}e^{-\frac{3}{2}\Omega}p_{\Omega}.
\end{align}
Note that $q>0$ and thus the range of the isotropic canonical variables is the open half-plane that admits the affine group of symmetry transformations (to be introduced later), an essential property used in our covariant quantization of the model. The Hamiltonian constraint (\ref{con}) is given by a sum of the isotropic and anisotropic parts,
\begin{align}\label{con2}\begin{split}
\mathrm{C}&=-\mathrm{C}_{iso}+\mathrm{C}_{ani},\\
\mathrm{C}_{iso}&=\frac{{\cal N}}{24}\left(\frac{9}{4}p^2+36q^{\frac{2}{3}}\right),\\
\mathrm{C}_{ani}&=\frac{{\cal N}}{24}\left(\frac{\mathbf{p}^2}{q^2}+36q^{\frac{2}{3}}V(\bsb)\right).\end{split}
\end{align}
The Hamilton equations  for ${\cal N}=24$ read:
\begin{align}\label{eqsymm}\begin{split}
\dot{q}=\frac{9}{2}p,&~~\dot{p} =-2\frac{\mathbf{p}^2}{q^3}+24q^{-\frac{1}{3}}[{V}(\bsb)-1],\\
\dot{\beta}_{\pm}=-2\frac{p_{\pm}}{q^2},&~~\dot{p}_{\pm}=36q^{\frac{2}{3}}\partial_{\pm}{V}(\bsb),
\end{split}
\end{align}
where $\partial_{\pm}:=\partial_{\beta_{\pm}}$. The above system of dynamical equations admits the following scaling symmetry:
\begin{align}\label{symm}\begin{split}
  t^{\prime}  &=\frac{t}{\delta^{1/2}},~~q^{\prime} = \frac{q}{\delta^{3/4}},~~p^{\prime} = \frac{p}{\delta^{1/4}},\\
   \beta^{\prime}_{\pm} &= \beta_{\pm},~~  p^{\prime}_{\pm} =\frac{p_{\pm} }{\delta},    \end{split}
\end{align}
{ where $\delta>0$ parametrizes the one-parameter group of scaling transformations acting on the variables of the dynamical model. They are symmetry transformations (labeled by $\delta$) that leave the dynamical equations invariant, that is, the primed variables defined in Eq. \eqref{symm} satisfy the same dynamical equations as the original variables in Eq. \eqref{eqsymm}. These transformations change some physical scale associated with the model, which one may expect to be no longer possible once some extra physical constant such as the Planck constant is introduced. This is what actually happens to our model as we later rigorously prove by applying the same transformations to the system of semiclassical dynamical equations and showing that in order for the equations to remain unchanged the semiclassical correction term must be rescaled as well, which in turn leads to genuinely different semiclassical model (as different size corrections correspond to different semiclassical models).  For more details, see Appendix B that is devoted to the scaling analysis of the semiclassical system. We note that these scalings can be viewed as transformations of large-universe solutions into small-universe solutions (or, vice versa), even into the ones that are smaller than the Planck scale at their recollapse. Hence it is natural to expect this symmetry to be broken at the quantum and semiclassical levels that must involve a new scale coming from the nonvanishing Planck constant.}

It is useful to express the dynamically most relevant geometric quantities in terms of the phase space variables:
\begin{align}\label{geoclas}\begin{split}
H&=\frac{p}{8q},~~R_{iso}=\frac{3}{2q^{\frac{4}{3}}},~~R_{ani}=-\frac{3V(\bsb)}{2q^{\frac{4}{3}}},\\
\sigma^2&=\frac{\mathbf{p}^2}{48q^4},\end{split}
\end{align}
where $H$, $R_{iso}$, $R_{ani}$ and $\sigma^2$ are respectively the Hubble rate, the isotropic intrinsic curvature, the anisotropic intrinsic curvature and the shear (squared). Upon rewriting in terms of them the constraint equation $\mathrm{C}_{iso}+\mathrm{C}_{ani}=0$ we obtain the generalized Friedmann equation:
\begin{align}\label{gfrw}
H^2=\frac{1}{6}\rho_r-\frac{1}{6}R_{iso}+\frac{1}{3}\sigma^2-\frac{1}{6}R_{ani},
\end{align}
where $\rho_r=M_r/q^{\frac{8}{3}}$ ($M_r$ is a constant) is the energy density of radiation that is added to the model.\footnote{The physical dimensions involved read: $[\kappa]=\frac{T^2}{ML}$, $[\frak{n}]=\frac{1}{L}$, $[\mathcal{V}_0]=L^3$, $[\omega^i]=L$, $[\mathcal{N}]=1$, $[q]=[\beta_{\pm}]=1$, $[p]=[p_{\pm}]=\frac{ML^2}{T}$, $[M_r]=\frac{ML^2}{T^2}$, where $M_r$ is the amount of the radiative energy in the universe when $V=\mathcal{V}_0$ (or, $q=1$).}

\section{Quantum model and its semiclassical portrait}
\label{quantsmod}
In the present section we explain the covariant integral quantization of the model. Our quantization method encompasses the ambiguities present in the quantization process and provides for them a convenient parametrization. This makes our analysis more robust. Moreover, our quantization method is naturally supplemented with a semiclassical framework that we shall use for the analysis of the mixmaster dynamics.

{ As was stressed in the article \cite{newref} and in the review paper \cite{berczugamal20}, our (integral)  quantization offers a wide choice of possibilities in establishing the map from a classical model to a quantum one for any physical quantity to be investigated. Quantizations based on operator valued integrals can be traced back to Weyl, Wigner, Klauder, Berezin, and many others. They have been recently developed in quantum cosmology studies with interesting results, like the regularization of singularities (see Refs \cite{berczugamal20,berczgamapie15A,berczgamapie15B,berczgama16B,berczgama16A,QC2014}). Naturally, from the physicist's viewpoint, the unique criterium of validity of one or a class of choices made among so many possibilities offered by the formalism is their agreement with measurements or observational data. The advantage of our approach with regard to other quantization procedures in use, like the canonical one, lies in the fact that there is a minimal amount of constraints imposed on the classical models, allowing for a phase space with singular boundaries, and the method yields a regularizing effect while  keeping a full probabilistic description in the subsequent semiclassical description, contrary to the Weyl-Wigner integral quantization.
}

\subsection{Covariant Weyl-Heisenberg integral quantization of functions on a  plane}
We consider a four-dimensional phase space $\R^4=\R^2\times\R^2$ made of two pairs of canonical variables, $(\beta_+,p_+)$ and $(\beta_-,p_-)$, and  define the integral quantization of a function $f(\mathbf{r}_{\pm})$ in the phase space $\mathbf{r}_{\pm}=(\beta_{\pm},p_{\pm})\in\mathbb{R}^2$ (we omit the index $_{\pm}$ in the sequel) as the following:
\begin{align}
f(\mathbf{r})\mapsto A_{f}:=\int_{\mathbb{R}^2} f(\mathbf{r})\mathcal{Q}(\mathbf{r})\frac{\ud^2\mathbf{r}}{2\pi},
\end{align}
where the $\mathcal{Q}(\mathbf{r})\in\mathcal{B}(\mathcal{H})$ are unit-trace operators  on $\mathcal{H}$, which resolve the identity
\begin{align}
\int_{\mathbb{R}^2} \mathcal{Q}(\mathbf{r})\frac{\ud^2\mathbf{r}}{2\pi}=\bu_{\mathcal{H}}.
\end{align}
From the arbitrariness of the choice of the origin of the phase space (in the absence of anisotropy potential), we make use of this translational symmetry, denoted by $\mathcal{T}: (\mathcal{T}(\mathbf{r}_0)f)(\mathbf{r})=f(\mathbf{r}-\mathbf{r}_0)$, and demand
\begin{align}
U(\mathbf{r}_0)A_{f}U(\mathbf{r}_0)^{\dagger}= A_{\mathcal{T}(\mathbf{r}_0)f},
\end{align}
where $\mathbf{r}\mapsto U(\mathbf{r})$ is a projective  unitary irreducible representation (UIR) of the group of translations in $\mathbb{R}^2$, equivalently a nontrivial UIR of the three-dimensional Weyl-Heisenberg group, 
\begin{align}
U(\mathbf{r})=e^{\ii(pQ-\beta P)},
\end{align}
where $Q$ and $P$ are the usual essentially self-adjoint position and momentum operators on the line with $[Q,P]=\ii \hbar \mathbbm{1}$. It turns out that any admissible family of operators $\mathcal{Q}(\mathbf{r})$ has the form
\begin{align}
\mathcal{Q}(\mathbf{r})=U(\mathbf{r})\mathcal{Q}_0U(\mathbf{r})^{\dagger},
\end{align}
where $\mathcal{Q}_0$ is a unit-trace operator. Thus, the choice of a quantization procedure is reduced to the choice of a single operator, $\mathcal{Q}_0$. Equivalently, one may use the weight function, $\Pi(\mathbf{r})$, which is defined via the Weyl-Heisenberg transform of $\mathcal{Q}_0$,
\begin{align}
\Pi(\mathbf{r}):=\mathrm{Tr}(U(-\mathbf{r})\mathcal{Q}_0)~~\Longrightarrow ~~\mathcal{Q}_0=\int_{\mathbb{R}^2}U(\mathbf{r})\Pi(\mathbf{r})\frac{\ud^2\mathbf{r}}{2\pi}~,
\end{align}
to determine the quantization procedure\footnote{We used the formula $\mathrm{Tr}(U(\mathbf{r}))=2\pi\delta(\mathbf{r})$.}. It is easy to see that $\mathrm{Tr}(\mathcal{Q}_0)=\Pi(0)$ and hence we must assume $\Pi(0)=1$. The weight $\Pi(\mathbf{r}_{\pm})$ defines the extent of coarse graining of the phase space $\mathbf{r}_{\pm}=(\beta_{\pm},p_{\pm})\in\mathbb{R}^2$. Notice that the standard canonical quantization (the Weyl quantization) is obtained for $\Pi(\mathbf{r})=1$, or equivalently for $\mathcal{Q}_0=2\mathsf{P}$, where $\mathsf{P}$ is the parity operator defined as $\mathsf{P}U(\mathbf{r})\mathsf{P}=U(-\mathbf{r})$.

\subsection{Semiclassical portraits}
Given a quantum operator $A_f$ corresponding to the observable $f$, we define the so-called quantum phase space portrait of the operator $A_f$ by making use of the same family of bounded unit-trace operators that we use for quantization,
\begin{align}
\reallywidecheck{f}(\mathbf{r})=\mathrm{Tr}(\mathcal{Q}(\mathbf{r})A_f)=\int_{\mathbb{R}^2} f(\mathbf{r}')\mathrm{Tr}(\mathcal{Q}(\mathbf{r})\mathcal{Q}(\mathbf{r}'))\frac{\ud^2\mathbf{r}'}{2\pi}~.
\end{align}
Actually, nothing save the simplicity prevents us from using another family of bounded unit-trace operators, say $\mathcal{R}(\mathbf{r})$, so that a more general quantum phase space portrait of  $A_f$ can be obtained,
\begin{align}
\reallywidecheck{f}(\mathbf{r})=\mathrm{Tr}(\mathcal{R}(\mathbf{r})A_f)=\int_{\mathbb{R}^2} f(\mathbf{r}')\mathrm{Tr}(\mathcal{R}(\mathbf{r})\mathcal{Q}(\mathbf{r}'))\frac{\ud^2\mathbf{r}'}{2\pi}~.
\end{align}
More tractable formulas are obtained when the weight function $\Pi(\mathbf{r})$ instead of the family of operators $\mathcal{Q}(\mathbf{r})$ (or, the defining $\mathcal{Q}_0$) is used. Let us define the symplectic Fourier transform of $f(\mathbf{r})$,
\begin{align}
\mathcal{F}[f](\mathbf{r})=\int_{\mathbb{R}^2}e^{-i\mathbf{r}\wedge\mathbf{r}'}f(\mathbf{r}')\frac{\ud^2\mathbf{r}'}{2\pi}~,
\end{align}
(where $\mathbf{r}\wedge\mathbf{r}'=\beta p'-p\beta'$). Now, it can be shown that 
\begin{align}
A_f=\int_{\mathbb{R}^2}U(\mathbf{r}) \mathcal{F}[f](-\mathbf{r})\Pi(\mathbf{r})\frac{\ud^2\mathbf{r}}{2\pi}
\end{align}
and
\begin{align}
\reallywidecheck{f}(\mathbf{r})=\int_{\mathbb{R}^2} \mathcal{F}[\Pi]*\mathcal{F}[\widetilde{\Pi}](\mathbf{r}'-\mathbf{r})f(\mathbf{r}')\frac{\ud^2\mathbf{r}'}{4\pi^2}~,
\end{align}
where $\widetilde{\Pi}(\mathbf{r})={\Pi}(-\mathbf{r})$. We shall study the mixmaster model at the semiclassical level, for which we need only the last formula, with the explicit knowledge of the operator $A_f$ becoming in fact unnecessary.

\subsection{Semiclassical portrait of the anisotropy}

\begin{center}
\begin{figure*}
\begin{tabular}{cc}
\includegraphics[width=0.38\textwidth]{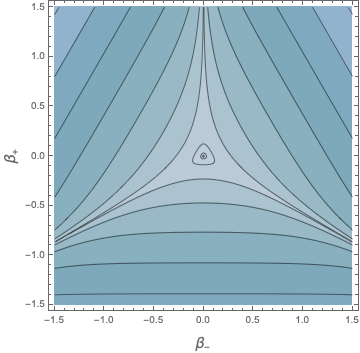}\hspace{1.5cm}
\includegraphics[width=0.38\textwidth]{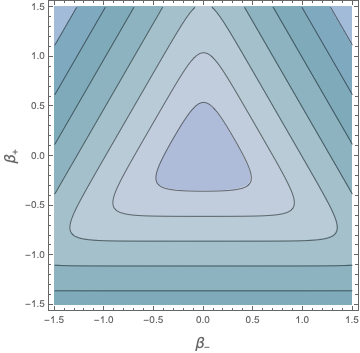}
\end{tabular}
\caption{The classical and semiclassical ($\tau_{\pm}=5=\sigma_{\pm}$) anisotropy potential. The classical potential comprises three narrowing channels with their bottoms asymptotically ($\beta_+\rightarrow\infty$ or $\beta_+\rightarrow -\infty$) approaching the zero value. In the semiclassical case the three channels become confined due to the semiclassical corrections as their bottoms raise indefinitely for $\beta_+\rightarrow\infty$ or $\beta_+\rightarrow -\infty$. } 
\label{1f}
\end{figure*}
\end{center}

Let us assume the following Gaussian weight function,
\begin{align}
\Pi(\beta,p)=e^{-\frac{\beta^2}{\sigma^2}}e^{-\frac{p^2}{\omega^2}},
\end{align}
where $\Pi(\beta,p)=\Pi(-\beta,-p)$ and $\Pi(0,0)=1$. The width parameters $\sigma$ and $\omega$ encode our degree of confidence in dealing with a given point in the phase space.  The symplectic Fourier transform of the weight reads,
\begin{align}
\mathcal{F}(\Pi)(\beta,p)=\frac{\sigma\omega}{2}e^{-\frac{1}{4}(\omega^2\beta^2+\sigma^2p^2)},
\end{align}
and their convolution reads
\begin{align}
\mathcal{F}(\Pi)\ast \mathcal{F}(\Pi)(\beta,p)=\frac{\pi\sigma\omega}{2} e^{-\frac{1}{8}(\omega^2\beta^2+\sigma^2p^2)}.
\end{align}
Hence, the lower symbol formula takes the form of regularizing Gaussian convolutions:
\begin{align}\label{regani}
\reallywidecheck{f}(\beta,p)=\int_{\mathbb{R}^2}\frac{\pi\sigma\omega}{2} e^{-\frac{1}{8}(\omega^2(\beta'-\beta)^2+(p'-p)^2\sigma^2)}f(\beta',p')\frac{\ud\beta'\ud p'}{4\pi^2}.
\end{align}
With this formula we easily find
\begin{align}
\reallywidecheck{(e^{-\alpha\beta})}=e^{\frac{4\alpha^2}{\omega^2}}e^{-\alpha\beta},~~\reallywidecheck{(p^2)}=p^2+\frac{8}{\sigma^2},
\end{align}
and the lower symbol of the anisotropy potential,
\begin{widetext}
\begin{align}\begin{split}
\reallywidecheck{V}(\beta_{\pm})&=\frac{1}{3}\left(D(4\sqrt{3},4)e^{4\sqrt{3}\beta_-+4\beta_+}+D(4\sqrt{3},4)e^{-4\sqrt{3}\beta_-+4\beta_+}+D(0,8)e^{-8\beta_+}\right)\\ &-\frac{2}{3}\left(D(2\sqrt{3},2)e^{-2\sqrt{3}\beta_--2\beta_+}+D(2\sqrt{3},2)e^{2\sqrt{3}\beta_--2\beta_+}+D(0,4)e^{4\beta_+}\right)+1,\end{split}
\end{align}
\end{widetext}
where the $D(x,y)=e^{\frac{4x^2}{\omega_-^2}}e^{\frac{4y^2}{\omega_+^2}}$ are regularization factors issued from our choice of the Gaussian weights. The classical and semiclassical anisotropy potentials are plotted in Fig. \ref{1f}.

\subsection{Quantization of the isotropy}
In analogy to the Weyl-Heisenberg group for the full plane, we adopt the so-called covariant affine group quantization (see for instance \cite{jpM2016} or a pedagogical presentation \cite{albegasca18}) of functions of the isotropic variables $(q,p)$ belonging to the half-plane $\R_+^\ast\times \R$. This quantization is consistent with the symmetry of the half-plane corresponding to the arbitrariness of the choice of the origin, namely $1$ for the scaling variable $q$ and $0$ for its conjugate momentum $p$. The respective phase space transformations form the affine group {Aff}$_+(\R)$ of the real line, that is,
\begin{align}
(q,p)\cdot (q',p')=\left(qq',\frac{p'}{q}+p\right)\in \R_+^\ast\times \R.
\end{align}
This group possesses two nonequivalent unitary irreducible representations (UIR) besides the trivial one. Both are square-integrable. One of them can be realized in the Hilbert space $\mathcal{H}=L^2(\R_+^*,\ud x)$. This UIR of {Aff}$_+(\R)$ acts on $\mathcal{H}$ as
\begin{align}
U(q,p)\psi(x)=e^{ipx}\frac{1}{\sqrt{q}}\psi(x/q).
\end{align}
For a normalized vector $\psi_0\in\mathcal{H}$, we introduce a continuous family of unit vectors as follows:
\begin{align}
(q,p)\mapsto \langle x|q,p\rangle := e^{ipx}\frac{1}{\sqrt{q}}\psi_0(x/q),
\end{align}
where $\psi_0$ is called the fiducial vector and the vectors $|q,p\rangle$ are called the affine coherent states (ACS). Let us define the constants $c_{\alpha}$ as
\begin{align}
c_{\alpha}:=\int_0^{\infty}|\psi_0(x)|^2\frac{\ud x}{x^{\alpha+2}}.
\end{align}
The resolution of unity by ACS is straightforward:
\begin{align}\label{resunit}
\int_{\R_+^\ast\times \R}\frac{\ud q\ud p}{2\pi c_{-1}}|q,p\rangle\langle q,p|=\bu_{\mathcal{H}},
\end{align}
provided that $c_{-1}<\infty$. Thanks to the above property of ACS, we define the quantization of the half-plane functions as:
\begin{align}\label{qascmap}
f(q,p)\mapsto A_f=\int_{\R_+^\ast\times \R}\frac{\ud q\ud p}{2\pi c_{-1}}f(q,p)|q,p\rangle\langle q,p|.
\end{align}
By construction, the quantization map is covariant with respect to the affine transformations:
\begin{align}
U(q',p')A_fU^\dagger(q',p')=A_{U(q',p')f},
\end{align}
where $[U(q',p')f](q,p)=f((q',p')^{-1}\cdot(q,p))$. 

\subsection{Semiclassical portrait of the isotropy}
The semiclassical portrait of a quantum operator $A_f$ is given by:
\begin{align}
\reallywidecheck{f}(q,p)=\langle q,p|A_f|q,p\rangle,
\end{align}
where the affine coherent states $|q,p\rangle$ are built with a fiducial vector that is in general different from the one used for quantization \eqref{qascmap}.

Combining the affine coherent state quantization with the semiclassical portrait of quantum operators yields the lower symbol of the phase space function $f$:
\begin{align}\label{aveqp}
\reallywidecheck{f}(q,p)=\int_{\R_+^\ast\times \R}\frac{\ud q'\ud p'}{2\pi c_{-1}}|\langle q,p|q',p'\rangle|^2f(q',p'),
\end{align}
which is the average of the function $f(q,p)$ with respect to the probability distribution $(q',p')\mapsto \dfrac{1}{2\pi c_{-1}}|\langle q,p|q',p'\rangle|^2$. As a fiducial vector we choose a family of unit vectors depending on the parameters $\xi$ and $\nu$:
\begin{align}
\psi_0(x)=\frac{1}{\sqrt{2x K_0(\nu)}}~e^{-\frac{\nu}{4}(\xi x+\frac{1}{\xi x})},
\end{align}
where $K_0$ is the modified Bessel function of the second kind. In order to simplify the calculations and ensure the commutation rule $[A_q,A_p]=1$ we fix $\xi=K_1(\nu)/K_2(\nu)$ (later for the sake of discussion of the classical limit we will restore the arbitrary $\xi$). We find the following lower symbols:
\begin{align}\begin{split}
\reallywidecheck{(p^2)}&=p^2+\frac{K(\nu)}{q^2},~~K(\nu)=\frac{K_1(\nu)^2\left(1+\nu\frac{K_0(\nu)}{K_1(\nu)}\right)}{4K_0(\nu)K_2(\nu)},\\
\reallywidecheck{(q^{\alpha})}&=Q_{\alpha}(\nu)q^{\alpha},~~Q_{\alpha}(\nu)=\frac{K_\alpha(\nu)K_{\alpha+1}(\nu)}{K_0(\nu)K_1(\nu)}.
\end{split}
\end{align}

\subsection{Semiclassical portrait of the total constraint}
\begin{center}
\begin{figure*}
\begin{tabular}{cc}
\includegraphics[width=0.3\textwidth]{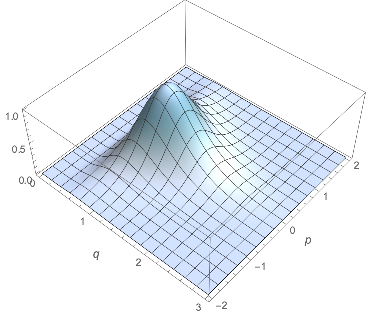}\hspace{1cm}
\includegraphics[width=0.3\textwidth]{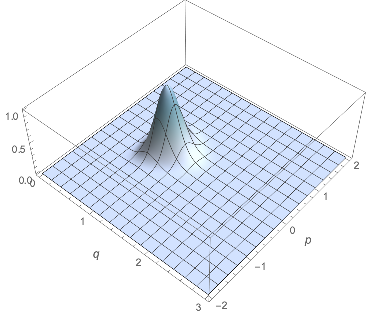}\hspace{1cm}
\includegraphics[width=0.3\textwidth]{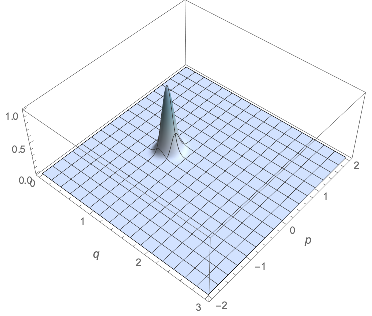}
\end{tabular}
\caption{The smearing probability distribution producing quantum corrections for $\nu=30,150,660$ (from left to right). For small values of $\nu$ our quantization procedure yields a very quantum system with large quantum uncertainties. On the other hand, for $\nu\rightarrow \infty$ our quantization procedure reproduces the exact classical system with vanishing uncertainties.} 
\label{2f}
\end{figure*}
\end{center}

The semiclassical portrait of the Hamiltonian constraint \eqref{con2} reads (for ${\cal N}=24$) as
\begin{align}\label{scconstr}\begin{split}
\reallywidecheck{\mathrm{C}}&=\frac{9}{4}\left(p^2+\frac{K(\nu)}{q^2}\right)-Q_{-2}(\nu)\frac{\mathbf{p}^2+\sum_{\pm}\frac{8}{\sigma_{\pm}^2}}{q^2}\\
&-36Q_{\frac{2}{3}}(\nu)q^{\frac{2}{3}}[\reallywidecheck{V}(\bsb)-1]\, .\end{split}
\end{align}
The quantum potential for the isotropic geometry $\propto q^{-2}$ is repulsive if and only if $\dfrac{9}{4}K(\nu)>Q_{-2}(\nu)\sum_{\pm}\dfrac{8}{\sigma_{\pm}^2}$, which we assume to hold in sequel. For convenience, we introduce
\begin{align}
K_{eff}(\nu,\sigma_{\pm}):=K(\nu)-\frac{32}{9}\sum_{\pm}\frac{Q_{-2}(\nu)}{\sigma_{\pm}^2}>0.
\end{align}
We derive from the semiclassical Hamiltonian constraint \eqref{scconstr} the following Hamilton equations:
\begin{align}
\label{Ham1}
\dot{q}&=\frac{9}{2}p\, , \\
\label{Ham2}\dot{p} &=\frac{9}{2}\frac{K_{eff}}{q^3}-2Q_{-2}\frac{\mathbf{p}^2}{q^3}+24Q_{\frac{2}{3}}q^{-\frac{1}{3}}[\reallywidecheck{V}(\bsb)-1]\,,\\
\label{Ham3}\dot{\beta}_{\pm}&=-2Q_{-2}\frac{p_{\pm}}{q^2}\,,\\
\label{Ham4}\dot{p}_{\pm}&=36Q_{\frac{2}{3}}q^{\frac{2}{3}}\partial_{\pm}\reallywidecheck{V}(\bsb)\,,
\end{align}
where
\begin{align}\begin{split}
\label{Q2Q23}
Q_{-2}&= Q_{-2}(\nu)=\frac{K_{2}(\nu)}{K_0(\nu)},\\
 Q_{\frac{2}{3}}&=Q_{\frac{2}{3}}(\nu)= \frac{K_{\frac{2}{3}}(\nu)K_{\frac{5}{3}}(\nu)}{K_0(\nu)K_1(\nu)}\, . \end{split}
\end{align}
We have thus obtained a semiclassical dynamical system in the full phase space $\R^\ast_+\times \R\times\R^4$ to be now examined. It involves six positive otherwise arbitrary quantization parameters: $\nu, \sigma_{\pm},\omega_{\pm}$ (degree of confidence...) and defines dynamical trajectories as a function of five initial conditions. We find the semiclassical model to be invariant under the following scalings:
\begin{align}\begin{split}
  t^{\prime}  &=\frac{t}{\delta^{1/2}},~~q^{\prime} = \frac{q}{\delta^{3/4}},~~p^{\prime} = \frac{p}{\delta^{1/4}},\\
   \beta^{\prime}_{\pm} &= \beta_{\pm},~~  p^{\prime}_{\pm} =\frac{p_{\pm} }{\delta},~~K_{eff}^{\prime}= \frac{K_{eff}}{\delta^2}.   \end{split}
\end{align}
We note that unlike the classical scale transformations \eqref{symm}, the above scalings involve $K_{eff}$, a nondynamical parameter that controls the quantum correction. This was to be expected as the quantization introduces a new scale into the system, i.e. the Planck scale, and thereby destroying the exact scaling symmetry present in the classical model. Nonetheless, if a solution to the semiclassical model with a fixed value of $K_{eff}$ is known, then respective solutions to the model for all the other values of $K_{eff}$ are also known. For a more detailed discussion of this scaling symmetry, see Appendixnonvanishing \ref{ASym}. 

It is straightforward to find the semiclassical versions of the geometric quantities \eqref{geoclas},
\begin{align}\label{geoquant}\begin{split}
\check{R}_{iso}&=\frac{3Q_{\frac{2}{3}}}{2q^{\frac{4}{3}}},~~\check{R}_{ani}=-\frac{3Q_{\frac{2}{3}}\check{V}(\bsb)}{2q^{\frac{4}{3}}},\\
\check{\sigma}^2&=\frac{Q_{-2}\mathbf{p}^2}{48q^4},~~\check{R}_Q=\frac{3K_{eff}}{32 q^4},\end{split}
\end{align}
as well as the semiclassical version of the generalized Friedmann equation \eqref{gfrw}:
\begin{align}\label{semFried}
H^2=\frac{1}{6}\rho_r-\frac{1}{6}\check{R}_{iso}+\frac{1}{3}\check{\sigma}^2-\frac{1}{6}\check{R}_{ani}-\frac{1}{6}\check{R}_Q,
\end{align}

We interpret the difference between the obtained semiclassical and the initial classical expressions to be the effect of quantum dispersion imposed on the geometry of the universe. The largest discrepancy between the classical and the semiclassical model is given by the repulsive potential $\dfrac{K_{eff}}{q^2}$ (or, equivalently, the quantum curvature $\check{R}_Q$). Another strong quantum feature is given by modifications to the anisotropy potential $\check{V}(\bsb)$. The remaining quantum features are introduced into the Hamilton equations \eqref{Ham1}-\eqref{Ham4} through the constants $Q_{-2}$ and $Q_{\frac{2}{3}}$. In particular, in Eq. \eqref{Ham3}, the constant $Q_{-2}$ alters the {\it classical} relation between the time derivative of the intrinsic three-metric $\beta_{\pm}$ and the extrinsic curvature $p_{\pm}$, which must hold in any {\it 4-d spacetime}. Therefore, with $Q_{-2}\neq 1$, a 4-d spacetime no longer exists\footnote{However, the requirement for the existence of the classical limit in this semiclassical model could be satisfied by renormalization of $\beta_{\pm}$ by the constant $Q_{-2}$.}. This comes from the fact that the momenta $(p,p_{\pm})$ do not commute with the three-geometry variables $(q,\beta_{\pm})$. The probability distribution smearing the isotropic $4$-geometry (introduced in Eq. \eqref{aveqp}) reads (after restoring the arbitrary parameter $\xi$)
\begin{align}
 \frac{1}{2\pi c_{-1}}\left|\frac{K_0\left(\nu\frac{q+q'}{2\sqrt{qq'}}\sqrt{1+\frac{4iqq'(p'-p)}{\nu\xi(q+q')}}\right)}{K_0(\nu)}\right|^2,
\end{align}
where the parameters $\nu$ and $\xi$ control the quantum dispersion induced by the affine coherent states. For $\xi=1/\nu$ and $\nu\rightarrow\infty$, the regularizing probability distribution converges to the Dirac delta $\delta(q-q')\delta(p-p')$ (in distributional sense) and all the physical quantities obtained from \eqref{aveqp} remain classical and satisfying the classical relations. However, the nonvanishing quantum uncertainty between $q$ and $p$ requires $\nu<\infty$, which produces a {\it smeared geometry}. In order to visualize the quantization process the probability distribution for the state $(q',p')=(1,0)$ for three different values of $\nu$ is plotted in Fig. \ref{2f}.

Analogously, the probability distribution associated with the lower symbol formula for anisotropic geometry \eqref{regani} is a Gaussian with the arbitrary parameters $\sigma_{\pm}$ and $\omega_{\pm}$. Taking  $\sigma_{\pm}\rightarrow\infty$ and $\omega_{\pm}\rightarrow\infty$ removes the quantum uncertainty between $\beta_{\pm}$ and $p_{\pm}$, the Gaussian probability distribution converges to the Dirac delta $\delta(\beta_{\pm}-\beta_{\pm}')\delta(p_{\pm}-p_{\pm}')$ and all the physical quantities obtained from \eqref{regani} retain their classical properties.


\section{semiclassical dynamics}
\label{semiclassdyn}
The anisotropy energy fuels the isotropic contraction and expansion. Moreover, the expansion and contraction can be fueled by various matter contributions. For simplicity, we keep only the radiation term in the Hamiltonian constraint, $-\dfrac{M_r}{q^{2/3}}$, where $M_r$ is a positive constant.

\subsection{Isotropic dynamics}\label{isocase}
\begin{figure*}[t]\centering
\begin{tabular}{cc}
\includegraphics[width=0.38\textwidth]{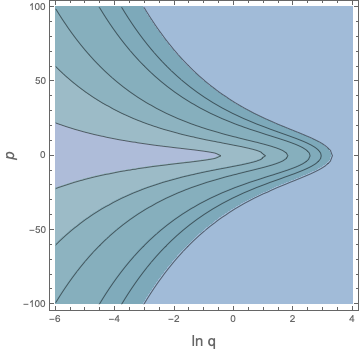}\hspace{1.5cm}
\includegraphics[width=0.38\textwidth]{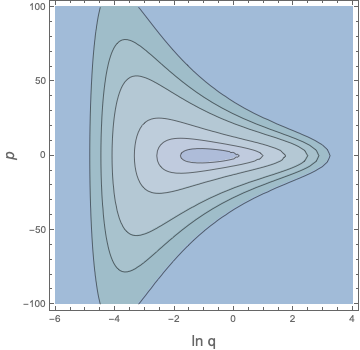}
\end{tabular}
\caption{ The isotropic classical-singular (on the left) and corresponding quantum-bouncing (on the right) solutions for various values of $M_r$ ($\nu=10$, $\sigma_{\pm}=5=\omega_{\pm}$).} 
\label{figure0}
\end{figure*}

Let us start by assuming perfectly spherical spatial sections with $\beta_{\pm}=0=p_{\pm}$. Then, the isotropic part of the constraint is the only nontrivially vanishing,
\begin{align}\label{isotropic}\begin{split}
\check{\mathrm{C}}_{iso}=\frac{9}{4}\left(p^2+\frac{K_{eff}}{q^2}\right)+36Q_{\frac{2}{3}}q^{\frac{2}{3}}-\frac{M_r}{q^{2/3}}\, .\end{split}
\end{align}
{ In this constraint equation, the first term is the expansion energy, the second is the repulsive potential, the third represents the positive intrinsic curvature and the last one is the (minus) energy of the fluid. With the shrinking volume (i.e., as $q$ decreases) the curvature term becomes negligible, while the repulsive potential grows until it perfectly balances the matter energy leading to the vanishing of the expansion energy ($p=0$). At this point the universe’s dynamics reverses and its volume starts to reexpand. This is a quantum bounce. On the other hand, with the growing volume (i.e., as $q$ increases) the repulsive term becomes negligible, while the curvature term grows until it perfectly balances the fluid energy leading to the vanishing of the expansion energy ($p=0$). At this point the universe’s dynamics reverses again and its volume starts to contract. This is a classical recollapse. See the right panel of Fig. \ref{figure0}.} 

For convenience, we introduce $L:=36Q_{\frac{2}{3}}$. Making use of the Eq. \eqref{Ham1} and the vanishing of the constraint \eqref{isotropic} we express the conformal time as a function of $q$,
\begin{align}\begin{split}
&\int\ud\eta=\int q^{-\frac{2}{3}}\ud t=\int \frac{\ud q}{q^{\frac{2}{3}}\dot{q}}=\frac{1}{4\sqrt{L}}\times\\
&\ln\left(2\sqrt{L}\sqrt{Lq^{\frac{8}{3}}+M_rq^{\frac{4}{3}}-\frac{9}{4}K_{eff}}+2Lq^{\frac{4}{3}}+M_r\right).\end{split}
\end{align}
The above relation is easily inverted if we neglect the intrinsic curvature, $L=0$, yielding the approximate solution
\begin{align}
q(\eta)=\left(4M_r\eta^2+\frac{9}{4}\frac{K_{eff}}{M_r}\right)^{\frac{3}{4}}~.
\end{align}
where the above approximation breaks down for large universes with non-negligible isotropic curvature.

The quantum bounce and the classical recollapse both occur when $p=0$, where the Hamiltonian constraint yields
\begin{align}
\frac{9}{4}K_{eff}+Lq^{\frac{8}{3}}-M_rq^{\frac{4}{3}}=0\, .
\end{align}
If we assume that at the quantum bounce the intrinsic curvature is negligible, i.e. $L=0$ as before, and at the classical recollapse the quantum repulsion is negligible, i.e. $K_{iso}=0$, then we find the minimal and maximal $q$ to read,
\begin{align}\label{qmin}
q_{min}=\left(\frac{9K_{eff}}{4M_r}\right)^{\frac{3}{4}},~~q_{max}=\left(\frac{M_r}{L}\right)^{\frac{3}{4}}.
\end{align}
Furthermore, the maximum $p=p_{max}$ occurs for $q$ such that
\begin{align}
\frac{\partial \check{\mathrm{C}}_{iso}}{\partial q}~\propto~\frac{9}{2}K_{eff}-\frac{2}{3}M_rq^{\frac{4}{3}}-\frac{2}{3}Lq^{\frac{8}{3}}=0,
\end{align}
which after neglecting $L$ gives $q=(27)^{\frac{1}{4}}q_{min}$, at which
\begin{align}
p_{max}= \frac{4}{3(27)^{\frac{1}{4}}\sqrt{2}}\frac{M_r^{3/4}}{K_{eff}^{1/4}}~.
\end{align}

A few { singular and corresponding bouncing} trajectories are plotted in Fig. \ref{figure0}. We note that in the isotropic case the phase of accelerated expansion is very brief and clearly insufficient from the point of view of the process of structure formation at a substantial range of cosmological scales. Indeed, combing the minimal value $q_{min}$ \eqref{qmin} with the value of $q$ at which the acceleration terminates (defined in Eq. \eqref{infl1} to be discussed later) we find that $\ln \dfrac{a_{end}}{a_{min}}=\ln \sqrt{2}$.

\begin{center}
\begin{figure*}
\begin{tabular}{cc}
\includegraphics[width=0.4\textwidth,height=0.4\textwidth]{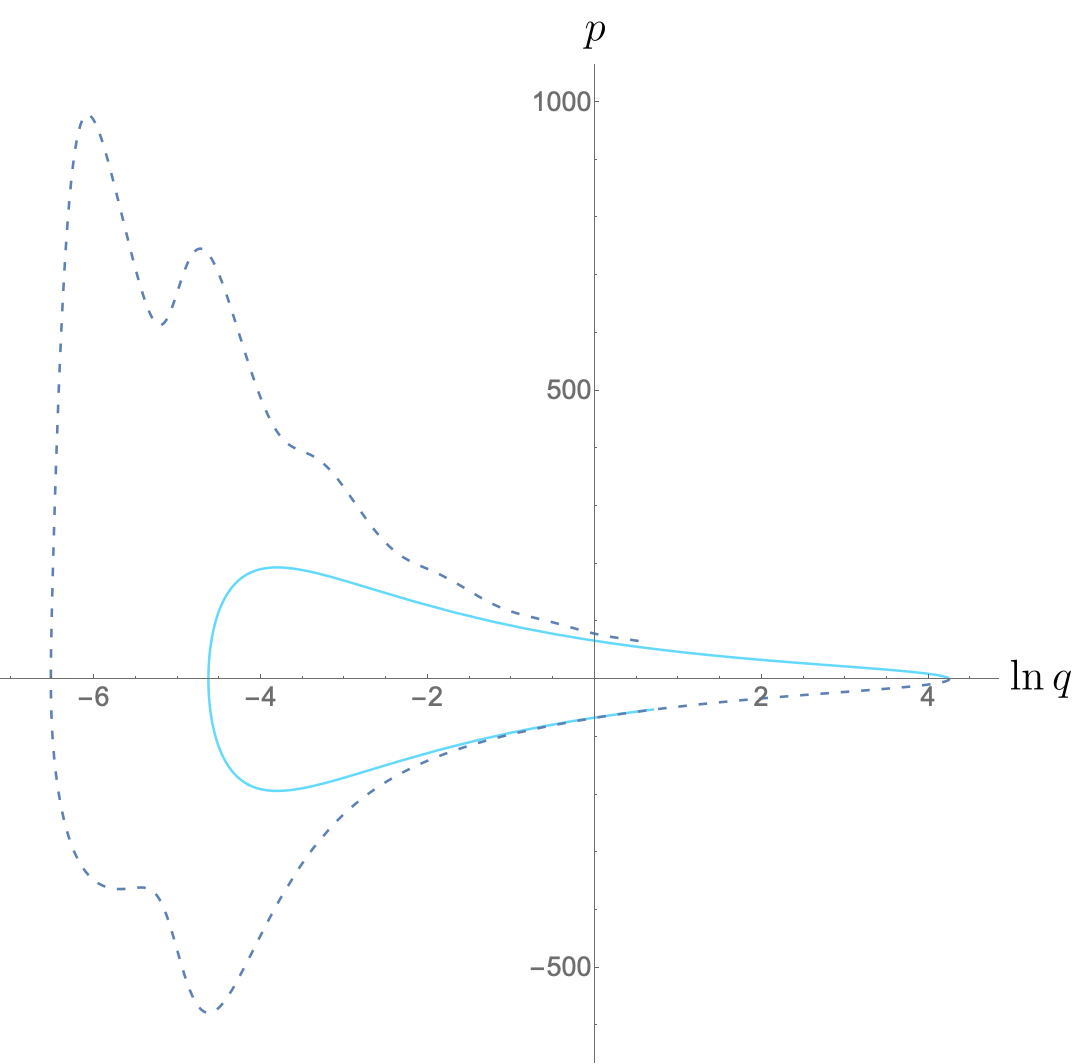}\hspace{1.5cm}
\includegraphics[width=0.4\textwidth,height=0.4\textwidth]{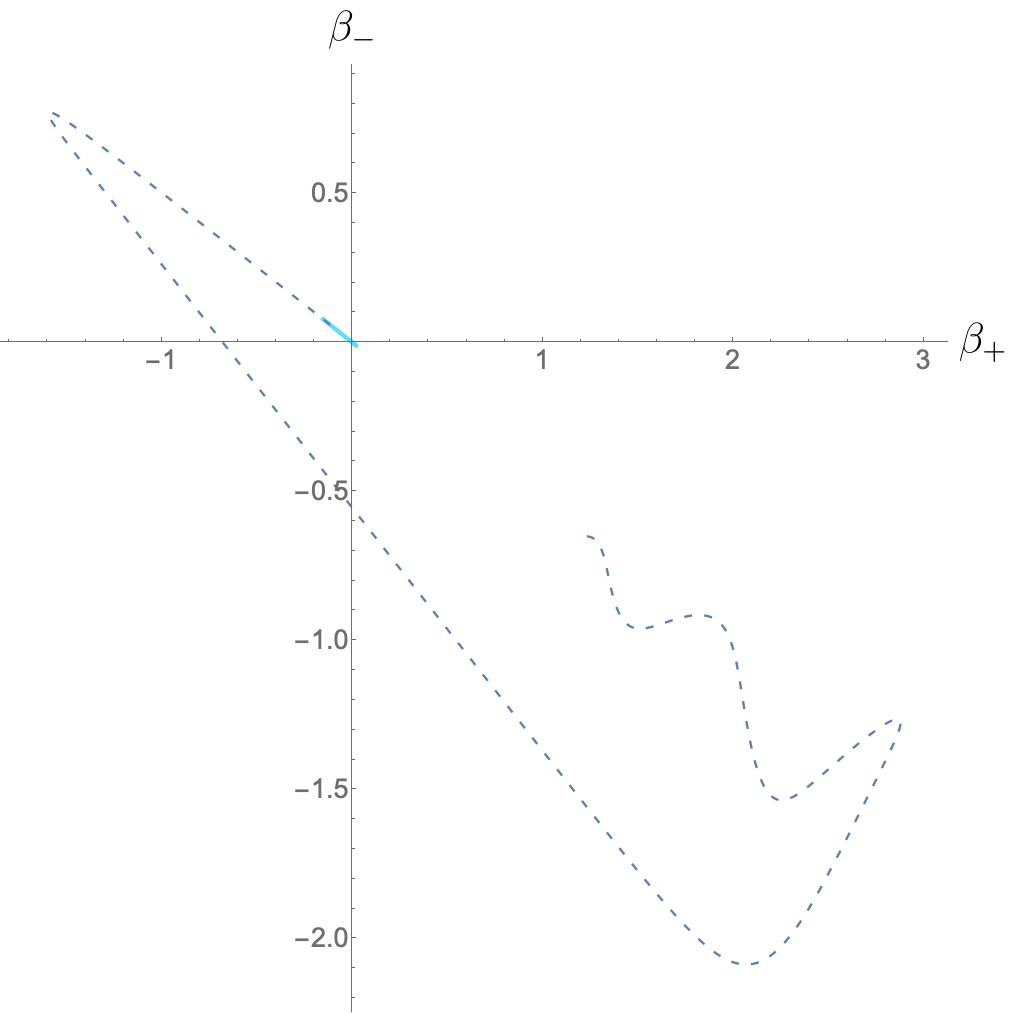}
\end{tabular}
\caption{Two cycles in the evolution of an anisotropic universe plotted in the $(q,p)$- and $(\beta_-,\beta_+)$-planes. The first and the second cycle are given by solid and dashed curves, respectively. Despite the fact that the first one is very isotropic and resembles the solutions of Fig. 3, the second one accumulates anisotropy on the approach to the bounce that now happens at a smaller volume. The dynamics around the second bounce is very asymmetric in the $(q,p)$-plane. The trajectory in the $(\beta_-,\beta_+)$-plane starts around the minimum where it remains for the first cycle. Then during the second cycle it moves to larger values of $\beta_\pm$ where it bounces off the potential walls producing oscillations. We set the following initial data:  $q=2.0$, $p=-52.6579$, $\beta_+=-0.01$, $\beta_-=0.005$, $p_+= 0.0$, $p_-= 0.0$. We set the parameters as follows: $\omega_\pm = 50$, $\sigma_{\pm} = 100$,  $\nu=37.5$  ($K_{eff}=9.255$),  $R=10^4$.} 
\label{sim1a}
\end{figure*}
\end{center}

\subsection{Anisotropic dynamics}

The anisotropy makes the dynamics of the universe too complex to be solved analytically. In order to reduce the complexity of the system it is common to employ the adiabatic approximation \cite{berczgamapie15A,berczgamapie15B}. In this approximation the complicated, oscillatory motions of the anisotropic variables are replaced with their energy averaged over many oscillations under the assumption that the value of the isotropic variable $q$ does not change significantly during this time. Moreover, the anisotropy potential that is responsible for the oscillations also requires an approximation such as the harmonic approximation or the steep-wall approximation. Unfortunately, these approximations have a rather restricted regime of applicability. Therefore, in the present work we choose to combine numerical computations with some analytical estimates.

In Fig \ref{sim1a} we present numerically integrated two cycles of a generic semiclassical mixmaster solution. As in the isotropic case, the anisotropic universe avoids the singularity through bounces. The quantum potential diminishes rapidly after the bounces and the anisotropy takes over the dynamics. Finally, the matter density exceeds the anisotropic energy density, and the standard Friedmann cosmology begins. Note that the two cycles (one given by the solid curve and the other by the dashed curve) are very different. The first one follows closely the isotropic solution and the $\beta_\pm$'s remian very small, whereas in the next cycle anisotropy develops as the universe contracts and the $\beta_\pm$'s start oscillating inside one of the channels. It results in an asymmetric bounce, leading to the destruction of the cosmic periodicity with each new cosmic cycle being different from the previous one.  One typically observes a few oscillations in the expansion rate right after the bounce. Moreover, a high rate of postbounce expansion can last for an extended period of time as seen from the behavior of the dynamical variable $p$. In \cite{berczgama16A} a similar behavior was observed and explained by the growth of the anisotropic energy triggered by the bounce. By the virtue of the Hamiltonian constraint, this newly produced anisotropic energy has to be balanced by the growth of the isotropic energy. During the bounces, the entire isotropic energy takes the form of the repulsive potential. As the universe starts to reexpand, the entire isotropic energy is transferred back to isotropic expansion. The observed dynamics points to the possibility for a phase of sustained accelerated postbounce expansion lasting for some $e$-folds. We shall investigate this issue in the following section.

Let us show now that a bounce exists in the generic semiclassical dynamics of the mixmaster universe. We note that the bounce must occur in the following subset of the constrained surface:
\begin{align}\label{p1}
p=0,~~\dot{p}=\frac{6K_{eff}}{q^3}-\frac{8Q_{-2}}{3}\frac{\mathbf{p}^2}{q^3}>0.
\end{align}
The above set of conditions defines a 4-dimensional subspace in the 5-dimensional constraint surface. A generic trajectory must pass through that region or even cross it infinitely many times. To see that in fact any trajectory should bounce let us follow the dynamics of $p$ along a typical trajectory in the constraint surface\footnote{The difference between $\dot{p}$ in Eqs \eqref{p1} and \eqref{p2} is vanishing at the constraint surface.}:
\begin{align}\label{p2}\begin{split}
\dot{p}&=\frac{\frac{9}{2}K_{eff}}{q^3}\\
&-\frac{2Q_{-2}}{q^3}\left\{\mathbf{p}^2-12\frac{Q_{\frac{2}{3}}}{Q_{-2}}q^{\frac{8}{3}}[\check{V}(\bsb)-1]\right\}.\end{split}
\end{align}
We note that the first and the third terms are positive while the second one negative, and their absolute values grow as the universe contracts with $q\rightarrow 0$ and $p<0$. As the anisotropic variables oscillate inside the potential walls the anisotropy energy (or, its part) is being transferred back and forth between the second (the kinetic) and the third (the potential) term. Initially, the sum of these two terms is negative and the first term is negligible as the universe is contracting more and more rapidly ($\dot{p}<0$). However, because of the oscillatory energy transfer, the absolute value of the two terms must grow slower than $q^{-3}$. Hence, down the line at some value of $q>0$ their sum must become dominated by the first term that is positive and grows as $q^{-3}$. As a result, for sufficiently small value of $q$ (provided that the bounce has not occurred before) the dynamics is sufficiently well approximated by 
\begin{align}
\dot{q}=\frac{9}{2}p,~~\dot{p}\approx\frac{\frac{9}{2}K_{eff}}{q^3},
\end{align}
leading to essentially the isotropic dynamics that we showed previously to be nonsingular. Hence, the bounce must eventually happen for any trajectory.
\begin{center}
\begin{figure*}
\begin{tabular}{cc}
\includegraphics[width=0.35\textwidth,height=0.35\textwidth]{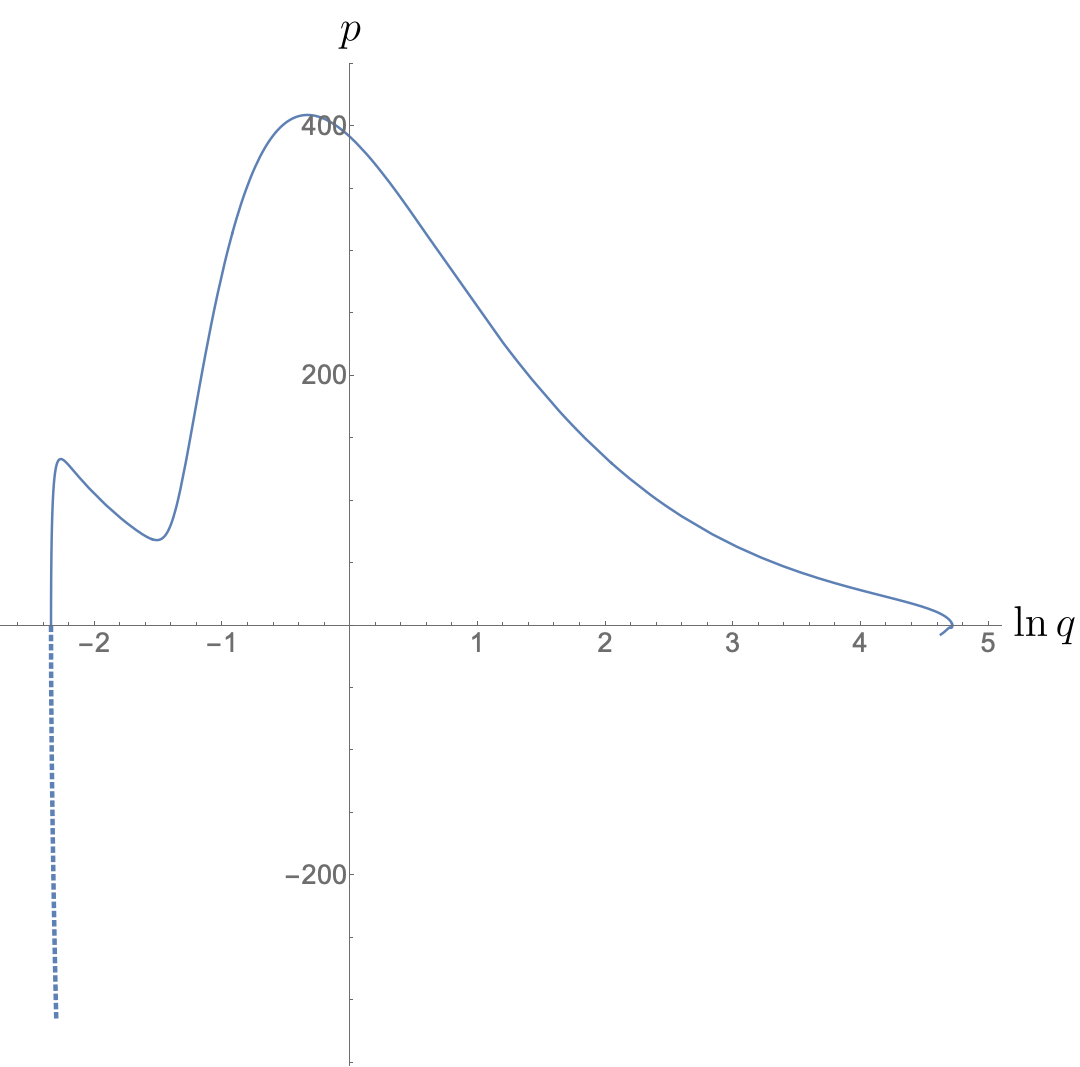}\hspace{1.5cm}
\includegraphics[width=0.35\textwidth,height=0.35\textwidth]{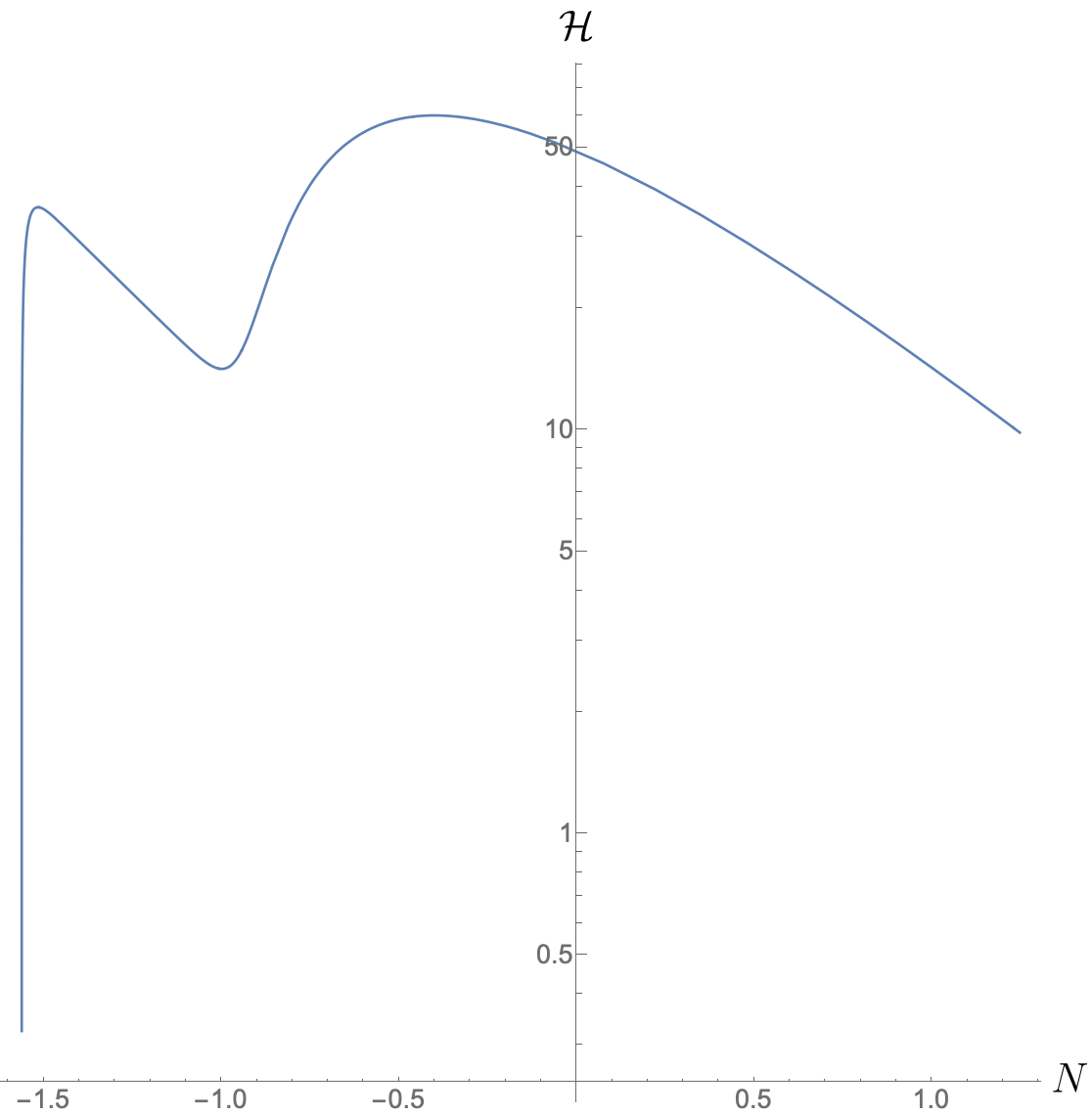}\\ 
\vspace{.7cm}\\
\includegraphics[width=0.35\textwidth,height=0.35\textwidth]{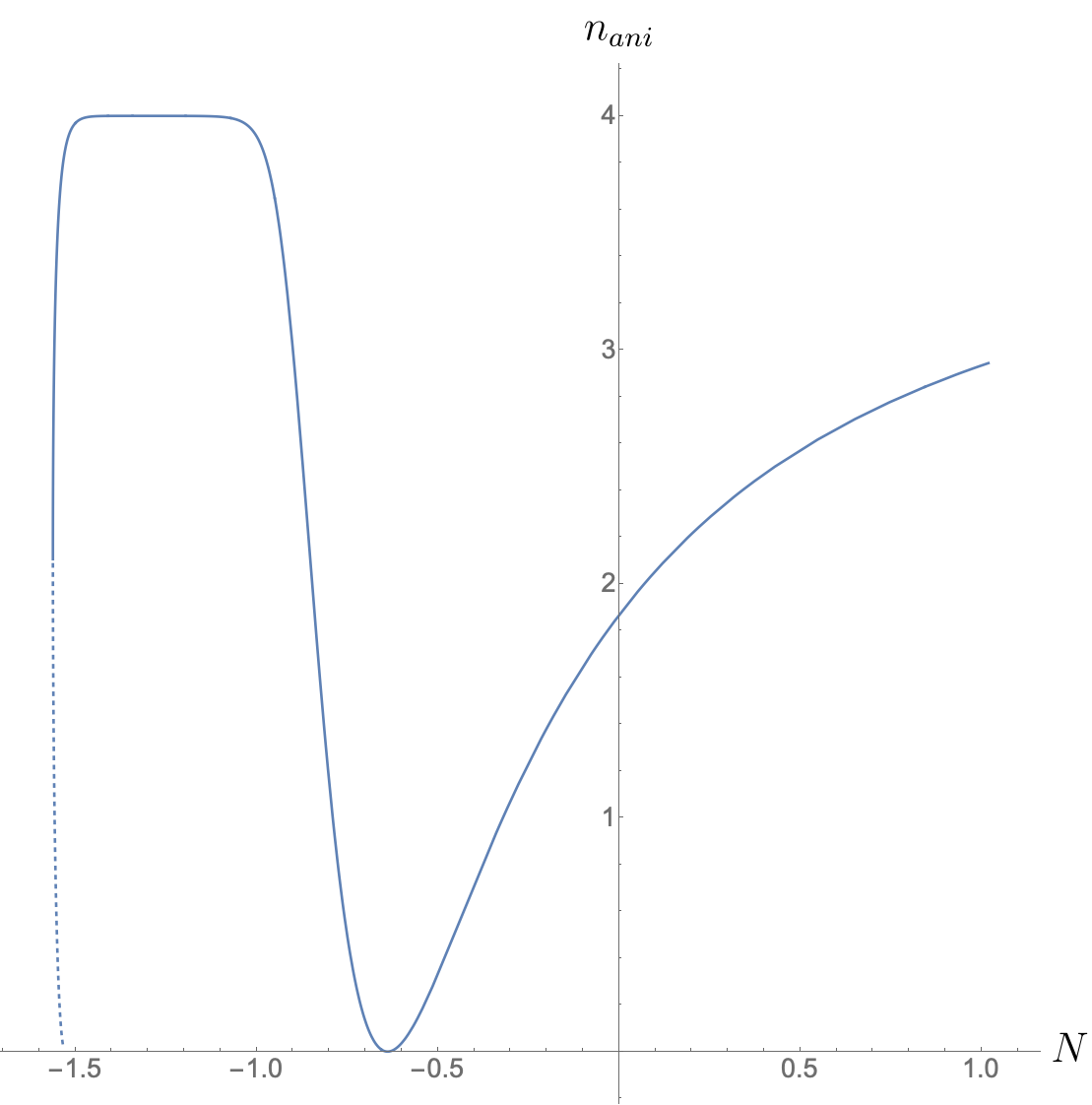}\hspace{1.5cm}
\includegraphics[width=0.35\textwidth,height=0.35\textwidth]{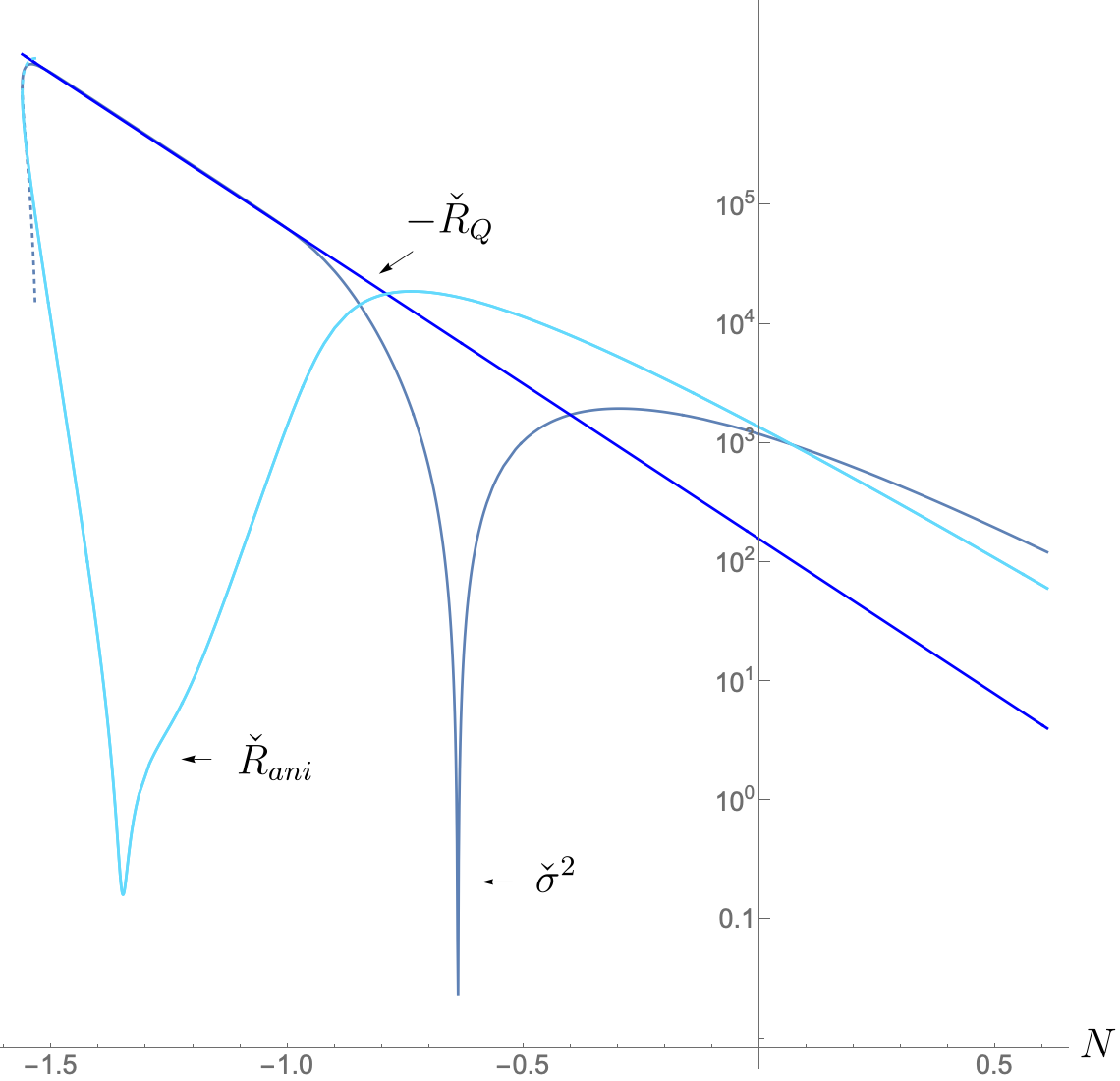}
\end{tabular}
\caption{A typical solution to the semiclassical dynamics of the mixmaster universe close to the bounce. In the $(q,p)$-plane the accelerated expansion is initially driven by the semiclassical correction as the universe bounces, then it ends and appears again driven by the anisotropic curvature. In the $(N,\mathcal{H})$-plane the inflationary dynamics is exhibited in the growth of the conformal Hubble rate $\mathcal{H}$, which takes place from around $N=-1$ to $N=-0.5$. In the $(N,n_{ani})$-plane the inflationary dynamics is reflected in the small values of $n_{ani}\approx 0$, which happens at the bounce and during the anisotropy-driven inflation (recall that $n_{ani}$ describes the behavior of the anisotropy energy, $\check{\rho}_{ani}\propto\frac{1}{a^{2+n_{ani}}}$). As the bottom-right panel shows, the dynamics is initially driven by the quantum curvature responsible for the bounce, then it is taken over by the anisotropy: first there is a lot of shear and little of anisotropic curvature so the dynamics is not inflationary. The inflationary dynamics begins once the energy of shear is transferred to the anisotropy potential which takes place around $N=-1$. We set the following initial data: $q=0.1$, $p= -312.895$, $\beta_+=0.0$, $\beta_-=-1.71$, $p_+= 0.0$, $p_-= 15.0$. We set the parameters as follows: $\omega_\pm = 56.23$, $\sigma_{\pm} = 100$,  $\nu=40008$  ($K_{eff}=10001.7$),  $R=10^2$.} 
\label{sim2a}
\end{figure*}
\end{center}

\section{Accelerated expansion}
\label{accelerated}
The quantum dynamics of the mixmaster universe is very rich and could, for instance, exhibit a prolonged phase of accelerated expansion during which the local structure inside the mixmaster universe is amplified in the same way as it happens for inflationary phase driven by a scalar field.

\subsection{General remarks}
The accelerated expansion takes place when $\ddot{a}>0$, or
\begin{align}\label{infl1}
\frac{\ud}{\ud\eta}\mathcal{H}>0,
\end{align}
where $\mathcal{H}=\acute{a}/a=\dot{a}$ is the conformal Hubble parameter\footnote{Differentiation with respect to cosmic and conformal time are denoted by `~$\dot{}$~' and `~$\acute{}$~', respectively.}. When the conformal Hubble horizon $\mathcal{H}^{-1}$ is shrinking, perturbation modes of fixed co-moving wavelengths leave the horizon and become amplified. It is often assumed that the span of wavelengths that exit the horizon during the inflationary phase is such that $k_{fin}/k_{ini}\gtrsim 10^{8}$. The growth in the number of wavelengths that cross the horizon reads:
\begin{align}
\frac{\ud k}{~k_{ini}}=\frac{\ud \mathcal{H}}{~\mathcal{H}_{ini}}=\frac{1}{~\mathcal{H}_{ini}}\frac{\ud \mathcal{H}}{\ud N}\ud N, 
\end{align}
where we have expressed the growth of the superhorizon scales as a function of the number of $e$-folds: $N=\ln ({a}/{a_{ini}})$. The range of scales that leave the horizon during a finite number of $e$-folds $\Delta N$ can be estimated from the initial state of the system via the Taylor expansion:
\begin{align}\label{taylor}\begin{split}
\frac{k_{fin}}{k_{ini}}&=\frac{1}{~\mathcal{H}}\frac{\ud \mathcal{H}}{\ud N}\Bigg|_{ini}\Delta N+\frac{1}{2}\frac{\ud}{\ud N}\left(\frac{1}{~\mathcal{H}}\frac{\ud \mathcal{H}}{\ud N}\right)\Bigg|_{ini}(\Delta N)^2\\
&+\mathcal{O}(\Delta N^2).\end{split}
\end{align}
If we assume that the second- (and any higher-) order term is much smaller than the first one, i.e., $\Big|\dfrac{(\Delta N)^2}{2}(\ln\mathcal{H})_{,NN}\Big|_{ini}\ll {k_{fin}}/{k_{ini}}$, then for most of the inflationary phase the Hubble horizon remains more or less constant and the phase lasts for $\Delta N=\dfrac{k_{fin}}{k_{ini}}\left(\dfrac{1}{~\mathcal{H}}\dfrac{\ud \mathcal{H}}{\ud N}\right)^{-1}\bigg|_{ini}$ of $e$-folds.

\subsection{Proper analysis}
If an inflationary phase occurs in the semiclassical mixmaster model it must be driven by either the quantum curvature $\frac{1}{6}\check{R}_Q$ or the anisotropy energy $\frac{1}{3}\check{\sigma}^2-\frac{1}{6}\check{R}_{ani}$, or a combination of both. We therefore neglect the radiation and the isotropic curvature and re-write the generalized Friedmann eq. \eqref{semFried} as 
\begin{align}\label{reFRW}\begin{split}
H^2=\frac{1}{6}\check{\rho}_{ani}-\frac{1}{6}\check{R}_Q,
\end{split}
\end{align}
where we introduced the notion of the anisotropy energy density $\frac{1}{6}\check{\rho}_{ani}=\frac{1}{3}\check{\sigma}^2-\frac{1}{6}\check{R}_{ani}$. 

We use the above equation to derive the Raychaudhuri equation with the expansion parameter replaced by the conformal Hubble parameter. Let us assume that at each moment of time the right-hand side terms are well approximated by power functions in the scale factor,
\begin{align}\label{rhoani}\begin{split}
\frac{1}{6}\check{\rho}_{ani}=\frac{\lambda_{ani}}{a^{n_{ani}+2}},~~\frac{1}{6}\check{R}_Q=\frac{\lambda_{Q}}{a^{6}},
\end{split}
\end{align}
($\lambda_{ani}>0$, $\lambda_{Q}>0$), i.e., their logarithms are approximately linear in the number of $e$-folds, $N \propto \ln a$. Note that the anisotropy effectively acts as a barotropic fluid with $\dfrac{\mathrm{p}}{\mathrm{\rho}}=w_{ani}=(n_{ani}-1)/3$. We find
\begin{align}
\frac{\ud}{\ud \eta}\mathcal{H}=\frac{a}{2}\frac{\ud}{\ud a}\mathcal{H}^2=-\frac{n_{ani}\lambda_{ani}}{2a^{n_{ani}}}+\frac{2\lambda_{Q}}{a^{4}}.
\end{align}
In order for the accelerated expansion to occur the condition \eqref{infl1} must hold, that is,
\begin{align}
0<\lambda_{Q}-\frac{n_{ani}\lambda_{ani}}{4a^{n_{ani}-4}},
\end{align}
which must be consistent with the Friedmann equation \eqref{reFRW},
\begin{align}\label{reFRWn}
0<\frac{\lambda_{ani}}{a^{n_{ani}-4}}-\lambda_{Q}.
\end{align}

Let us first assume that $\lambda_{Q}= 0$, that is, the influence of the quantum curvature on the expansion is negligible. It is possible only if $n_{ani}< 0$. This behavior coincides with the barotropic fluid behavior with $w_{ani}< -\frac{1}{3}$. We will see below that this behavior of the anisotropy energy density is impossible neither in the classical nor in the semiclassical model.

Since we neglect the radiation it is impossible to neglect anisotropy by putting $\lambda_{ani}= 0$, i.e., it has to be present in the expanding universe if the condition \eqref{reFRWn} is to hold.

The remaining possibility is that both the anisotropy and the quantum curvature are important in the expanding universe. In this case the above conditions are combined into ($n_{ani}>0$):
\begin{align}
n_{ani}\lambda_{Q}<\frac{n_{ani}\lambda_{ani}}{a^{n_{ani}-4}}<4\lambda_{Q},
\end{align}
from which we see immediately that $0<n_{ani}<4$. Upon dividing the above inequality by $n_{ani}\lambda_{Q}$ and fixing $\frac{\lambda_{ani}}{\lambda_{Q}}e^{N_i(4-n_{ani})}=1$, $\frac{\lambda_{ani}}{\lambda_{Q}}e^{N_f(4-n_{ani})}=\frac{4}{n_{ani}}$, we obtain
\begin{align}\label{nani}
e^{\Delta N(4-n_{ani})}=\frac{4}{n_{ani}},
\end{align}
where $\Delta N=N_f-N_i$. One may verify that there are two solutions to the above equation for $n_{ani}$ if $\Delta N>\dfrac{1}{4}$. Since the accelerated expansion must occur for around $\Delta N=20$ $e$-folds, $n_{ani}$ can be very small and close to $n_{ani}=4e^{-4\Delta N}$. This behavior coincides with the barotropic fluid behavior for the barotropic parameter $w_{ani}\approx -\frac{1}{3}$, which, as we show below, cannot last sufficiently long to yield a robust inflationary phase. Another solution is $n_{ani}=4$, which lies in the closure of the admissible values but does not belong to them. Hence, we exclude this solution. 

Note that we may also interpret Eq. \eqref{nani} as yielding the number of $e$-folds for a given value of $n_{ani}$. Since $n_{ani}< 4$ we conclude that the lower bound for the number of $e$-folds reads $\Delta N=0.25$. This lower bound implies that anisotropy can in fact reduce the duration of the inflationary phase with respect to the isotropic radiation-filled universe, for which the number of $e$-folds $\Delta N=\ln\sqrt{2}\approx 0.347\ll 20$ is clearly above the found lower bound (see Sec. \ref{isocase}), though still much too small for the inflationary scenario to be relevant in that case.

Let us now inspect the equations of motion for anisotropy. We use the analogy between scalar fields in isotropic universe and the anisotropy variables. Upon dividing the anisotropic part of the semiclassical Hamiltonian \eqref{scconstr} by $-36Q_{\frac{2}{3}}q^{\frac{2}{3}}$ (or, by setting $1/\mathcal{N}:=-36Q_{\frac{2}{3}}q^{\frac{2}{3}}$) it acquires the following form:
\begin{align}\label{hamb}
H_{ani}=\frac{\mathbf{p}^2}{2m}+\reallywidecheck{V}(\bsb),
\end{align}
where the mass $m(q)=18Q_{\frac{2}{3}}q^{\frac{8}{3}}/Q_{-2}$ depends on the size of the universe in such a way that $m$ grows as the universe expands. The equation of motion for $\beta_{\pm}$ reads
\begin{align}\label{eomb}
\ddot{\beta}_{\pm}=-\frac{1}{m}\reallywidecheck{V}_{,\beta_{\pm}}-\frac{\dot{m}}{m}\dot{\beta}_{\pm}.
\end{align}
This dynamics is conservative only when $\dot{q}=0$. However, $\dot{q}>0$ as the universe expands, and hence the energy $H_{ani}$ may only decrease. Given that the anisotropy energy density at each moment of time behaves as a power function of the scale factor (see Eq. \eqref{rhoani}), we have
\begin{align}
\check{\rho}_{ani}\propto \frac{H_{ani}}{a^2}\propto \frac{1}{a^{n_{ani}+2}},\end{align}
with $n_{ani}>0$ as was to be shown. Upon inspecting the Hamiltonian \eqref{hamb} we clearly see that the kinetic energy scales as $a^{-4}$, whereas the potential energy is independent of the scale factor. Hence, we conclude that $0<n_{ani}<4$.

In order to reproduce the inflationary dynamics, we must have $n_{ani}=4e^{-4\Delta N}$, which is positive and very small for $\Delta N= 20$ $e$-folds. This requires the dynamics to be dominated by the anisotropy potential with a negligible kinetic energy $\dot{\beta}_{\pm}\approx 0$. In other words, the relative change of the potential during that number of $e$-folds must be very small. We find
\begin{align}
\ud \reallywidecheck{V}=\reallywidecheck{V}_{,\pm}\ud\beta_{\pm}=\reallywidecheck{V}_{,\pm}\frac{\dot{\beta}_{\pm}}{H}\Delta N=-\reallywidecheck{V}_{,\pm}p_{\pm}\frac{Q_{-2}\Delta N}{12 q^2 H},
\end{align}
where $H$ is the Hubble rate. Let us assume $\ud \reallywidecheck{V}=-\dfrac{\mathbf{p}^2}{2m}\bigg|_{fin}$ (at the end of inflation) and combine it with the last relation to obtain
\begin{align}
\ud \reallywidecheck{V}=\left[\reallywidecheck{V}_{,\pm}\frac{\sqrt{2m}Q_{-2}\Delta N}{12 q^2 H}\right]^2.
\end{align}
Thus, the condition $\ud \reallywidecheck{V}/\reallywidecheck{V}\ll1$ implies
\begin{align}
\frac{\reallywidecheck{V}_{,\pm}}{\reallywidecheck{V}}\ll \frac{2 \mathcal{H}/\sqrt{\reallywidecheck{V}}}{\sqrt{Q_{\frac{2}{3}}/Q_{-2}}\Delta N},
\end{align}
where the conformal Hubble rate reads roughly $\mathcal{H}\approx\frac{Q_{\frac{2}{3}}}{2}\sqrt{\reallywidecheck{V}}$ (by the virtue of the constraint equation) yielding
\begin{align}
\frac{\reallywidecheck{V}_{,\pm}}{\reallywidecheck{V}}\ll \frac{\sqrt{Q_{\frac{2}{3}}Q_{-2}}}{\Delta N}.
\end{align}
It is easy to see that $2<\frac{|\reallywidecheck{V}_{,\pm}|}{|\reallywidecheck{V}|}<8$ except close to the point of origin $\bsb=0$, where the potential $\reallywidecheck{V}$ has the minimum. We see that neither classical nor semiclassical potential can satisfy the above requirement and hence a sustained inflationary phase is excluded from this model. It is the exponential character of $\reallywidecheck{V}$ that disallows anisotropy-driven inflation. A typical postbounce evolution is plotted in Fig. \ref{sim2a}. 

At this point it is interesting to note that the inflationary phase might occur in the harmonic approximation of the anisotropy potential as
\begin{align}
\frac{|V_{,\pm}|}{V}=\frac{|2\bsb|}{\bsb^2}=\frac{2}{|\bsb|},
\end{align}
can be smaller than any value provided that the particle is placed sufficiently far away from the point of origin $\bsb=0$. This explains our previous result obtained in a full quantum framework in \cite{berczgama16A}, where the harmonic approximation to the anisotropy potential was used. 

\begin{center}
\begin{figure*}
\begin{tabular}{cc}
\includegraphics[width=0.4\textwidth]{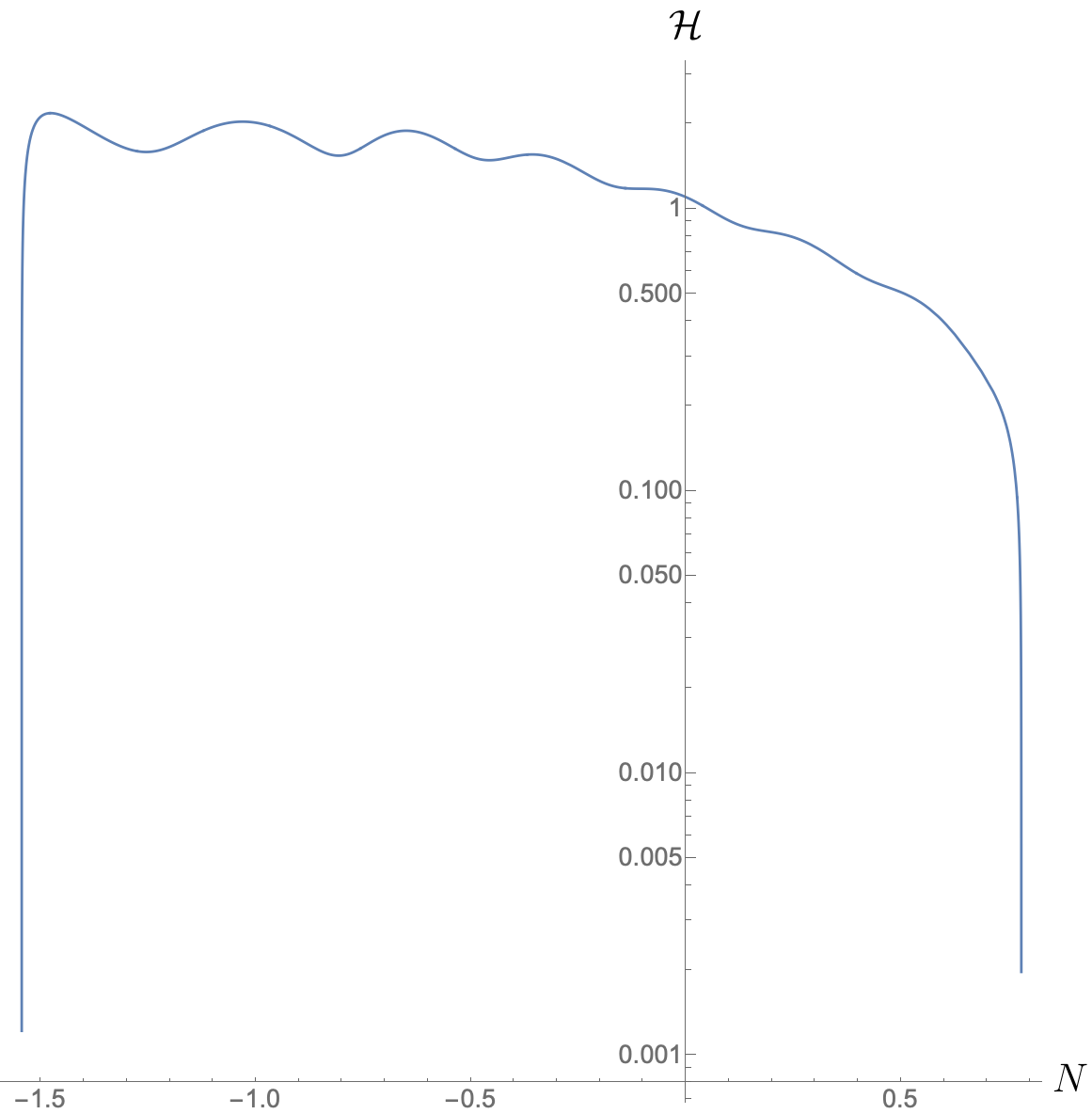}\hspace{1.5cm}
\includegraphics[width=0.4\textwidth]{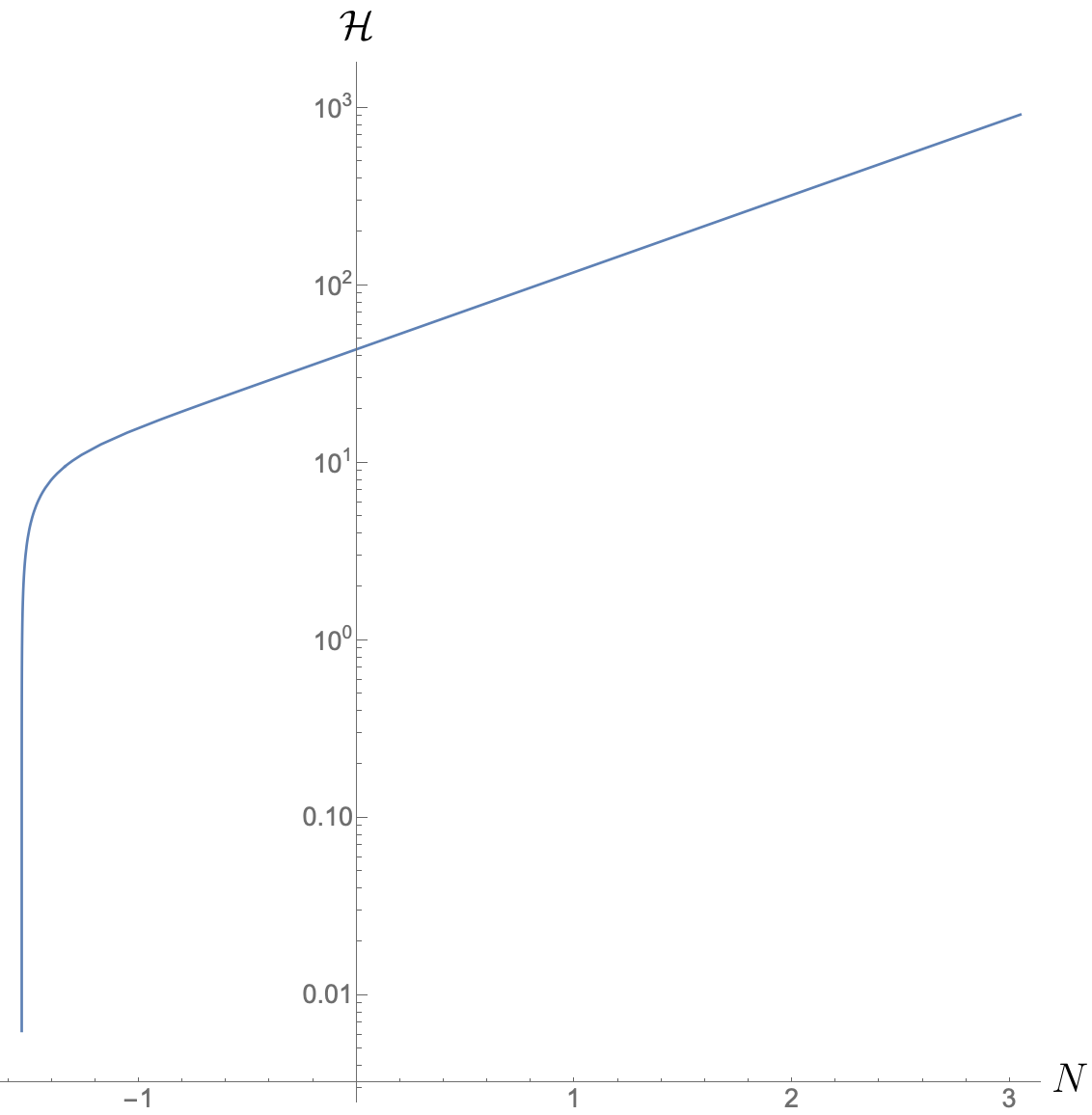}
\end{tabular}
\caption{The evolution of the conformal Hubble rate in the anisotropy- and inflaton-dominated universes. On the left we see a robust exponential growth of the conformal Hubble rate generated by inflaton in a quadratic potential. On the left we see how the anisotropic curvature (just after the bounce ends) increases the conformal Hubble rate only by little and in an oscillatory manner (due to the oscillating anisotropic deformations $\beta_\pm$). We set the following initial data: $q=0.1$, $p= -5.139$, $\beta_+=0.579$, $\beta_-=0.748$, $p_+= 0.175$, $p_-= 0.15$, $\phi=2615121.8$, $\pi_{\phi}=0.01$. We set the parameters as follows: $\omega_\pm = 177.83$, $\sigma_{\pm}^2 = 0.04738$,  $\nu=350$  ($K_{eff}=11.9083$),  $R=1.0$, the mass of inflaton $m_\phi=0.0000578$.} 
\label{sim3a}
\end{figure*}
\end{center}

It is instructive to compare the anisotropy Hamiltonian to that of a minimally coupled scalar field in a closed universe. The scalar field Hamiltonian constraint reads:
\begin{align}
\mathrm{C}_{\phi}=\mathcal{N}\left(\frac{1}{2q^{2}}\pi_{\phi}^2+q^2V(\phi)\right),
\end{align}
which can be brought to the form of Eq. \eqref{hamb} by setting $\mathcal{N}:=1/q^2$:
\begin{align}
H_{\phi}=\frac{1}{2m'}\pi_{\phi}^2+V(\phi),
\end{align}
where $m'(q)=q^4$. Now, we see clearly that the key difference lies in the respective masses $m(q)$ and  $m'(q)$. The energy density of the scalar field now reads
\begin{align}
{\rho}_{\phi}\propto {H}_{\phi}\propto \frac{1}{a^{n_{\phi}}},\end{align}
where $6>n_{\phi}>0$ following from the same reasoning as before. We clearly see that because of the minimal coupling the requirements for the inflationary potential are much milder than for anisotropy potential. Furthermore, given a complete (or, almost complete) freedom in proposing the inflationary potential, one may choose the harmonic one that easily produces the desired accelerated expansion. The numerical comparison of inflation- and anisotropy-driven dynamics is given in Fig. \ref{sim3a}.

\section{Conclusions}
\label{disc}
In this work we investigated whether a quantum anisotropic primordial universe can spontaneously induce a phase of inflationary dynamics. We first derived a very generic quantum model of mixmaster universe via integral covariant quantization and coherent states methods. Thanks to these methods we were able to cover a wide range possible quantization ambiguities and semiclassical frameworks parametrized by a set of constants. Then using the equations of motion we found the reasons for why anisotropic universe, neither classical nor quantum, cannot induce a sustained inflationary phase in the early universe. In order to state these reasons clearly we compared the anisotropic model to the single-field inflationary model.

Both the anisotropy and the inflaton energy are declining as the universe expands. However, a minimally coupled field can produce effective pressure with $w_{\phi}\in (-1,1)$ while anisotropy produces effective pressure with $w_{ani}\in (-\frac{1}{3},1)$. It is well known that $w_{eff}<-\frac{1}{3}$ is required in order for accelerated expansion to take place. Therefore, pure anisotropy fails to induce inflation. Nevertheless, if one adds to the system a quantum correction in the form of repulsive potential, then any contribution, including anisotropy, can induce inflation for  $w_{ani}\approx -1/3$. Thus, in principle, anisotropy could induce a sustained inflationary phase if its potential allowed for it. The crucial property is that in order for its effective pressure to remain minimal the relative change of the potential in the configuration space of a given model should be very small. For the anisotropy potential this is however impossible because the potential is fixed by general relativity to be exponential. The inflationary potential does not have this limitation and could be, e.g., quadratic. We note that even quantization of the anisotropy potential does not change its exponential character.

Our analysis was semiclassical and perhaps going to a full quantum description (in particular, of anisotropy) could change the character of the solution. Furthermore, if we included the backreaction from quantum perturbations, the anisotropy potential could perhaps acquire large corrections allowing for sustained inflationary phase. Having that said we believe that neither of these two options is very probable. Therefore, we propose another cosmological scenario in which anisotropy plays a key role in the generation of primordial structure. It is a bouncing cosmology in which the generation starts in the contracting phase and then is smoothly transferred through a bounce to the expanding phase. It is well known that given a single cosmic fluid with $w=\frac{p}{\rho}$ being its equation of state, bouncing cosmology yields the spectral index $n_s-1=6(1+w)/(1+3w)$ or $n_s-1=12w/(1+3w)$ \cite{Malk2023}, which is always blue-tilted (for $w>0$) contrary to the observed one that is slightly red-tilted \cite{PLANCK}. The addition of anisotropy would produce an effective cosmological fluid which could lead to the primordial perturbations with the correct spectrum. In this scenario, one could expect, there is more anisotropy in the contracting phase than in the expanding. We postpone the investigation of the details of this proposal to a future paper.

\begin{acknowledgments}
P.M. and J.C.M. acknowledge the support of the National Science Centre
(NCN, Poland) under the research grant 2018/30/E/ST2/00370.
\end{acknowledgments}

\appendix
\section{Isotropic dynamics}
For negligible $L$ we find that the amount of time needed for the universe to bounce back to the  same volume $q^2$ reads,{\scriptsize
\begin{align}\nonumber\begin{split}
\Delta\eta&=\frac{1}{\sqrt{R}}\sqrt{q^{\frac{4}{3}}-\frac{9}{4}\frac{K_{iso}}{R}},\\
\Delta t&=\frac{1}{2\sqrt{R}}\left[q^{\frac{2}{3}}\sqrt{q^{\frac{4}{3}}-\frac{9K_{iso}}{4R}}+\frac{9K_{iso}}{4R}\ln\left(\frac{q^{\frac{2}{3}}+\sqrt{q^{\frac{4}{3}}-\frac{9K_{iso}}{4R}}}{\sqrt{\frac{9K_{iso}}{4R}}}\right)\right],\end{split}
\end{align}}
which is useful for numerical integrations.

\section{Symmetry analysis  of the semiclassical dynamics}\label{ASym}
Let us rescale the four variables $t$, $q$, $p_{\pm}$ in the following way:
\begin{equation}
\label{scaling1}
t= \gamma t^{\prime}\, , \quad q= \lambda q^{\prime}\, , \quad p_{\pm}= \delta p_{\pm}^{\prime}\, ,  
\end{equation}
while keeping $\beta_{\pm}= \beta^{\prime}_{\pm}$. 
The dynamical system in the primed variables reads:
\begin{align}
\label{Ham1p}
\dot{q^{\prime}}&=\frac{9}{2}p^{\prime}\, , \quad \mbox{with}\quad p^{\prime}= \frac{\gamma}{\lambda}p\, , \\
\label{Ham2p}\begin{split}\dot{p^{\prime}} &=\frac{9}{2}\frac{\gamma^2}{\lambda^4}\frac{K}{{q^{\prime}}^3}-2\frac{\gamma^2\delta^2}{\lambda^4}Q_{-2}\frac{{p^{\prime}}_{\pm}^2+\frac{8}{(\delta\sigma_{\pm})^2}}{{q^{\prime}}^3}\\
&+24\frac{\gamma^2}{\lambda^{4/3}}Q_{\frac{2}{3}}{q^{\prime}}^{-\frac{1}{3}}[\reallywidecheck{V}(\bsb^{\prime})-1]\,,\end{split}\\
\label{Ham3p}\dot{\beta^{\prime}}_{\pm}&=-2\frac{\delta\gamma}{\lambda^2}Q_{-2}\frac{p^{\prime}_{\pm}}{{q^{\prime}}^2}\,,\\
\label{Ham4p}\dot{p^{\prime}}_{\pm}&=36\frac{\gamma\lambda^{2/3}}{\delta}Q_{\frac{2}{3}}{q^{\prime}}^{\frac{2}{3}}\partial_{\pm}\reallywidecheck{V}(\bsb^{\prime})\,,
\end{align}
The original system is almost regained if additionally the following relations are imposed:
\begin{align}
\label{relscal1}
  \sigma^{\prime}_{\pm}  &= \delta\sigma_{\pm}\,,     \\
\label{relscal2}  Q^{\prime}_{\frac{2}{3}}  &= \frac{\gamma^2}{\lambda^{4/3}} Q_{\frac{2}{3}}   = \frac{\gamma\lambda^{2/3}}{\delta}Q_{\frac{2}{3}}\, , \\
 \label{relscal3} Q^{\prime}_{-2}  &= \frac{\gamma^2\delta^2}{\lambda^{4}} Q_{-2}   = \frac{\gamma\delta}{\lambda^2}Q_{-2}\, , \\
\label{relscal4}  K_{eff}^{\prime}&= \frac{\gamma^2}{\lambda^4}K_{eff}\,. 
\end{align} 
From Eq. \eqref{relscal3} we have $\lambda= \sqrt{\gamma\delta}$, and thus $Q^{\prime}_{-2}=Q_{-2}$. Then Eqs \eqref{relscal2} are compatible and yield 
$Q^{\prime}_{\frac{2}{3}}  = \left(\dfrac{\gamma^2}{\delta}\right)^{2/3} Q_{\frac{2}{3}} $. Hence, $Q^{\prime}_{\frac{2}{3}}=Q_{\frac{2}{3}}$ implies  $\gamma= \sqrt{\delta}$.
We finally obtain the following primed dynamical system:
\begin{align}
\label{Ham1pf}
\dot{q^{\prime}}&=\frac{9}{2}p^{\prime}\, ,  \\
\label{Ham2pf}\dot{p^{\prime}} &=\frac{9}{2}\frac{K^{\prime}}{{q^{\prime}}^3}-2Q_{-2}\frac{{p^{\prime}}_{\pm}^2+\frac{8}{(\sigma^{\prime}_{\pm})^2}}{{q^{\prime}}^3}\\ \nonumber
&+24Q_{\frac{2}{3}}{q^{\prime}}^{-\frac{1}{3}}[\reallywidecheck{V}(\bsb^{\prime})-1]\,,\\
\label{Ham3pf}\dot{\beta^{\prime}}_{\pm}&=-2Q_{-2}\frac{p^{\prime}_{\pm}}{{q^{\prime}}^2}\,,\\
\label{Ham4pf}\dot{p^{\prime}}_{\pm}&=36Q_{\frac{2}{3}}{q^{\prime}}^{\frac{2}{3}}\partial_{\pm}\reallywidecheck{V}(\bsb^{\prime})\,,
\end{align}
where the scaling relations between primed and non-primed variables and parameters involve only powers of $\delta$,
\begin{align}
\label{scaledelta}
  t^{\prime}  &=\frac{t}{\delta^{1/2}}\, ,    \\
  q^{\prime} &= \frac{q}{\delta^{3/4}}\,,\\
   p^{\prime} &= \frac{p}{\delta^{1/4}}\,,\\
   \beta^{\prime}_{\pm} &= \beta_{\pm} \,,\\
   p^{\prime}_{\pm} & =\frac{p_{\pm} }{\delta}\, , \\
    \sigma^{\prime}_{\pm}  &= \delta\sigma_{\pm}\,,     \\
    K^{\prime}&= \frac{K}{\delta^2}\, . 
\end{align}

\end{document}